\documentclass{article}

\usepackage{arxiv}

\usepackage[utf8]{inputenc} 
\usepackage[T1]{fontenc}    
\usepackage{hyperref}       
\usepackage{url}            
\usepackage{booktabs}       
\usepackage{amsfonts}       
\usepackage{nicefrac}       
\usepackage{microtype}      
\usepackage{lipsum}
\usepackage{mathrsfs}  
\usepackage{graphicx}
\graphicspath{ {./images/} }
\usepackage{amsmath}
\usepackage{amssymb}
\usepackage{mathtools}
\usepackage{genyoungtabtikz}
\usepackage[title]{appendix}
\usepackage{float}
\usepackage{natbib}
\usepackage{comment}
\setcitestyle{authoryear, open={(},close={)}}

\newtheorem{proposition}{Proposition}
\newtheorem{corollary}{Corollary}
\usepackage[title]{appendix}

\newtheorem{theorem}{Theorem}
\newcommand{\rnew}{r^{\text{new}}}
\newcommand{\rold}{r^{\text{old}}}
\title{A new representation of Ewens-Pitman's partition structure and its characterization via Riordan array sums}
\numberwithin{equation}{section}

\author{
 Jan Greve\\
  Institute for Statistics and Mathematics\\
  WU Vienna\\
  Welthandelspl 1, 1020 Vienna\\
  \texttt{jgreve@wu.ac.at}
}

\begin{document}
\maketitle
\begin{abstract}
Ewens-Pitman's partition structure arises as a system of sampling consistent probability distributions on set partitions induced by the Pitman-Yor process. It is widely used in statistical applications, particularly in species sampling models in Bayesian nonparametrics. Drawing references from the area of representation theory of the infinite symmetric group, we view Ewens-Pitman's partition structure as an example of a non-extreme harmonic function on a branching graph, specifically, the Kingman graph. Taking this perspective enables us to obtain combinatorial and algebraic constructions of this distribution using the interpolation polynomial approach proposed by Borodin and Olshanski (\textit{The Electronic Journal of Combinatorics}, \textbf{7}, 2000). We provide a new explicit representation of Ewens-Pitman's partition structure using modern umbral interpolation based on Sheffer polynomial sequences. In addition, we show that a certain type of marginals of this distribution can be computed using weighted row sums of a Riordan array. In this way, we show that some summary statistics and estimators derived from Ewens-Pitman's partition structure can be obtained using methods of generating functions. This approach simplifies otherwise cumbersome calculations of these quantities often involving various special combinatorial functions. In addition, it has the added benefit of being amenable to symbolic computation.
\end{abstract}


\section{Introduction}
Probability distributions on set and integer partitions appear in various areas in statistics such as clustering \citep{fraley2002model,mccullagh2008many,miller2018mixture,fruhwirth2021generalized,greve2022spying}, estimation of unseen species \citep{good1956number,efron1976estimating,mao2004predicting,lijoi2007bayesian,lee2013defining,balocchi2024bayesian}, natural language processing \citep{teh2004sharing,teh-2006-hierarchical,teh2010hierarchical} and statistical disclosure control \citep{samuels1998bayesian,hoshino2001applying,favaro2021bayesian}. While applications to broader areas of science include examples in physics \citep{higgs1995frequency}, rumor spreading \citep{taga1959stochastic,bartholomew1971stochastic}, trawling \citep{sibuya2014prediction} and many more. Interestingly, foundational works in the theory of random partitions can be traced back not only to probability theory \citep{ferguson1973bayesian,blackwell1973ferguson,aldous1996probability,pitman1997two} but also to the area of population genetics \citep{ewens1972sampling,kingman1978representation,kingman1978random}. In particular, \cite{kingman1978representation} coined the term \textit{partition structure} which refers to a system of distributions on set partitions subject to the sampling consistency condition that enables distributions of different sample sizes to be connected through a recursion. For statistical problems with an emphasis on extrapolation from the sample rather than simply describing the data at hand, this condition is indispensable. Species sampling models \citep{ishwaran2003generalized} in Bayesian nonparametrics deal precisely with this problem of extrapolation and have therefore embraced models such as the Dirichlet process \citep{ferguson1973bayesian} and the Pitman-Yor process \citep{pitman1997two} whose finite-dimensional distributions on partitions retain sampling consistency. 

The main subject of this paper is Ewens-Pitman's partition structure indexed by two parameters $0\leq \alpha<1$ and $\theta> -\alpha$, which is also called Ewens-Pitman's sampling formula. Notable historical developments of Ewens-Pitman's partition structure started with the work by \cite{kingman1978random} who established the bijection between the two-parameter Poission-Dirichlet measure $\text{PD}(\alpha,\theta)$ on the infinite-dimensional Kingman simplex and Ewens-Pitman's partition structure. Later on, Ewens-Pitman's partition structure also became known as the induced random partition from the Pitman-Yor process \citep{pitman1997two}. For a comprehensive review of the historical development of partition structures and applications including the aforementioned literature, see \cite{10.1214/15-STS529}.

Ewens-Pitman's partition structure is commonly viewed as a distribution on partitions of the set of integer-valued labels $\{1,\ldots,n\}$ invariant under the permutations of these labels. Therefore, one can also consider the state space of this discrete distribution to be the set of all partitions of $n\in\mathbb{N}$. Consider partitions of the set $\{1,2,\ldots,n\}$ of size $n$ with $l\in\{1,\ldots,n\}$ parts with counts within each part arranged as $\lambda_1\geq\ldots\geq\lambda_l>0$ in descending order. The combined probability of these equivalent set partitions of class $\lambda_1\geq\ldots\geq\lambda_l>0$ following Ewens-Pitman's partition structure is given by:
\begin{align}
    M_n(\lambda) = \binom{n}{\lambda_1,\ldots,\lambda_l}\varphi_{\theta,\alpha}(\lambda)\label{eq:EP_form}
\end{align}
for an integer partition $\lambda = (\lambda_1,\ldots,\lambda_l), \sum_{i=1}^l\lambda_i = n$. The function $\varphi_{\theta,\alpha}$ has a form
\begin{align}
    \varphi_{\theta,\alpha}(\lambda) = \frac{\theta^{l\uparrow\alpha}}{\theta^{n\uparrow}\prod_{j=1}^nr_j(\lambda)!}\prod_{i=1}^l (1-\alpha)^{\lambda_i-1\uparrow}\label{eq:EP_harmonic}
\end{align}
where $r_j(\lambda)=\sum_{i=1}^l 1\{\lambda_i = j\}$ is the number of parts with size $j$ for an integer partition $\lambda$ while $x^{k\uparrow a}$ for real numbers $x$ and $a$, and a nonnegative integer $k$ is defined as 
\begin{align}
    x^{k\uparrow a} = \begin{cases}
        1 &\text{for } k = 0\\
        x(x+a)\cdots(x+(k-1)a) & \text{for } k=1,2,\ldots,
    \end{cases}
\end{align}
with $x^{k\uparrow}:= x^{k\uparrow 1}$. The definition \eqref{eq:EP_form} of Ewens-Pitman's partition structure $M_n(\lambda)$ as a product of a multinomial coefficient and a function $\varphi_{\theta,\alpha}$ of form \eqref{eq:EP_harmonic} is widely adopted in the literature of representation theory of the infinite symmetric group (see \citealt{borodin2017representations} and references therein for an overview) and surrounding areas. In particular, \citet[Theorem 4.1]{borodin1999harmonic} showed that $\varphi_{\theta,\alpha}$ of form \eqref{eq:EP_harmonic} is an exmample of a \textit{non-extreme harmonic function} on a \textit{branching graph}, specifically the Kingman graph and introduced a method to construct this distribution called \textit{interpolation polynomial} approach. Other literature in this area concerning Ewens-Pitman's partition structure are \cite{kerov1989combinatorial}, \cite{kerov2006coherent} and \cite{petrov2009two} to name a few. In addition, \cite{gnedin2006exchangeable} widely cited among Bayesian nonparametrics literature can also be closely aligned to this area.

\subsection{Main results}

In this work, we show that the interpolation polynomial approach \citep{borodin1999harmonic} to construct $\varphi_{\theta,\alpha}$ on the Kingman graph  results in a generalized form of the Newton method for polynomial interpolation called (modern) \textit{umbral interpolation} \citep{costabile2025umbral}. As a result, we obtain a new explicit representation of $\varphi_{\theta,\alpha}$ for a given integer partition $\lambda$ of size $n$ with $l$ parts as follows:
\begin{align}
\varphi_{\theta,\alpha}(\lambda)&=\frac{(-1)^n}{\theta^{n\uparrow}}\sum_{i=0}^{l}m^*_\lambda(\underbrace{\alpha,\ldots,\alpha}_{l+i})(-1)^i\frac{(\theta+l\alpha)^{l+1\uparrow\alpha}}{(\theta+(l+i)\alpha)\alpha^{l}(l-i)!i!}\\
m^*_\lambda(\underbrace{\alpha,\ldots,\alpha}_{l+i}) &= \frac{(l+i)^{l\downarrow}}{\prod_{j=1}^nr_j(\lambda)!}\prod_{i=1}^l\alpha^{\lambda_i\downarrow}
\end{align}
where $x^{k\downarrow a}$ for real numbers $x$ and $a$, and a nonnegative integer $k$ is defined as 
\begin{align}
    x^{k\downarrow a} = \begin{cases}
        1 &\text{for } k = 0\\
        x(x-a)\cdots(x-(k-1)a) & \text{for } k=1,2,\ldots,
    \end{cases}
\end{align}
with $x^{k\downarrow}:= x^{k\downarrow 1}$. The function $m^*_\lambda$ is called the \textit{factorial monomial symmetric function}. Hence, Ewens-Pitman's partition structure is written as:
\begin{align}
    M_n(\lambda) = \frac{n!(-1)^n}{\prod_{i=1}^l\lambda_i!\theta^{n\uparrow}}\sum_{i=0}^{l}m^*_\lambda(\underbrace{\alpha,\ldots,\alpha}_{l+i})(-1)^i\frac{(\theta+l\alpha)^{l+1\uparrow\alpha}}{(\theta+(l+i)\alpha)\alpha^{l}(l-i)!i!}\label{result:representationMn}.
\end{align}
This representation explicitly shows the decomposition between $m^*_\lambda$, a term that depends on partition sizes parameterized solely with $\alpha$, and the term $(\theta+l\alpha)^{l+1\uparrow\alpha}/((\theta+(l+i)\alpha)\alpha^{l}(l-i)!i!)$ which depends on the partition length $l$ parameterized by both $\alpha$ and $\theta$. From this result, one can deduce the important characteristics of Ewens-Pitman's partition structure that partition length $l$ is the sufficient statistics for $\theta$ while conditional on $l$, partition sizes in $\lambda$ is the sufficient statistics for $\alpha$ \citep{pitman2006combinatorial,10.1214/15-STS529,mano2018partitions}.

Additionally, by adopting the representation \eqref{result:representationMn}, we show that the marginal of $M_n(\lambda)$ taken with respect to all interger partitions $\lambda$ of size $n$ with length $l$ is given as:
\begin{align}
    \sum_{\substack{\lambda\in\mathscr{P}_n\\\text{s.t. }l(\lambda) =l}} M_n(\lambda) = \frac{S_{n,l}(-1,-\alpha,0)}{\theta^{n\uparrow}}(-\alpha)^l\Big[\frac{t^l}{l!}\Big]\Big(\frac{-1}{t-1}\Big)\Big(1-\frac{t}{t-1}\Big)^{-\theta/\alpha-l}\label{result:marginal}
\end{align}
where $\mathscr{P}_n$ is the set of all partitions of $n$, $l(\lambda)$ is the length of a given integer partition $\lambda$, $S_{n,l}(-1,-\alpha,0)$ is the special case of a generalized Stirling number $S_{n,l}(a,b,c)$ given in \cite{hsu1998unified} and $[t^l/l!]$ is an operator that takes the coefficient of the term $t^l/l!$. Importantly, we show that the term
$$
\Big[\frac{t^l}{l!}\Big]\Big(\frac{-1}{t-1}\Big)\Big(1-\frac{t}{t-1}\Big)^{-\theta/\alpha-l}
$$
in \eqref{result:marginal} is a linear combination of elements in the $n$-th row of an Exponential \textit{Riordan array} \citep{shapiro2022riordan} to which one can apply the \textit{fundamental theorem of Riordan arrays} (FTRA) to obtain the above form as a coefficient of a power series. We use this result to obtain alternative derivations for several summary statistics and estimators involving Ewens-Pitman's partition structure given in Bayesian nonparametric literature such as \cite{lijoi2008bayesian},\cite{cerquetti2013marginals} and \cite{favaro2013conditional}, as well as an earlier work by \cite{yamato2000moments}. Our derivation consists mostly of symbolic methods based on (exponential) generating functions.

\subsection{Organization of the paper}

The organization of the paper is as follows. Section \ref{sec:KerovVershikStuff} is a self-contained introduction to the intersection between probability theory and representation theory of the infinite symmetric group which covers important concepts such as branching graphs and the harmonic function on these graphs. Then, it posits that Ewens-Pitman's partition structure is an example of a non-extreme harmonic function on the Kingman graph and introduces the interpolation polynomial approach in \cite{borodin1999harmonic} to construct this distribution. Finally, a brief coverage of the Newton polynomial interpolation to facilitate this construction is given with an example. In Section \ref{sec:RiordanArrayShefferSeq}, a deeper look at the Newton polynomial interpolation as an example of a modern umbral interpolation via Sheffer sequences is given. This results in, to the best of our knowledge, a new explicit representation of Ewens-Pitman's partition structure. In addition, Riordan arrays are introduced with an important operation called the fundamental theorem of Riordan arrays (FTRA). Riordan arrays form a group isomorphic to the group formed by Sheffer sequences. We use this isomorphism to show that a certain marginal of Ewens-Pitman's partition structure has an FTRA representation. Section \ref{Sec:Estimators} shows one application of the new representation of Ewens-Pitman's partition structure and its marginals to derive summary statistics and estimators known in the literature, under a unified use of the symbolic methods of generating functions. While none of the results in this section are new, the derivation is new and amenable to symbolic computer algebra software implementations.  Section \ref{sec:Conclusion} concludes with some discussions.

\section{Representation Theory of the Infinite Symmetric Group and Construction of Ewens-Pitman’s Partition Structure}\label{sec:KerovVershikStuff}
One of the areas in representation theory of the infinite symmetric group $S(\infty)$ is the theory on $S(\infty)$-invariant measures, i.e., measures that are invariant under the permutation of indices in the population $\mathbb{N}$. The representation of $S(\infty)$-invariant measures developed in this area has a direct correspondence with de Finetti's theorem on exchangeable sequences of random variables. Methods to construct Ewens-Pitman's partition structure which is proportional to the exchangeable partition probability function (EPPF, \citealt{pitman1995exchangeable}) have therefore also been developed in this area. In this section, we introduce basic notations and concepts in representation theory of the infinite symmetric group and utilize the method of interpolation polynomials proposed in \cite{borodin1999harmonic} to construct Ewens-Pitman's partition structure.
\subsection{Integer partitions and Young diagrams}\label{sec:partitions_youngdiagram}
A partition $\lambda$ of a positive integer $n$ of length $l$ belonging to the set of all integer partitions $\mathscr{P}_n$ can be uniquely identified using several notations. In this work, we utilize three notations interchangeably. The first notation treats $\lambda$ as an infinite sequence of nonnegative integers with finitely many nonzero parts:
\begin{align}
    \lambda = (\lambda_1,\lambda_2,\ldots,\lambda_l,0,0,\ldots),\qquad \lambda_1\geq\lambda_2\geq\ldots\geq\lambda_l > 0.\nonumber
\end{align}
The second notation is the first notation without the zeros on the tail. This results in a sequence of length $l$ which depends on $\lambda$ and hence we denote the length of $\lambda$ in this notation as $l(\lambda):=l$. In addition, we use $|\lambda|$ as the notation for the sum of the integers of the partition $\lambda$, resulting in $|
\lambda|=\sum_{i=0}^\infty \lambda_i = \sum_{i=0}^l \lambda_i = n$ and call it the size of the partition $\lambda$. The final notation, commonly referred to as the \textit{frequency} notation counts the number of parts with a value of $j$ as $r_j(\lambda) = \sum_{i=1}^l 1\{\lambda_i = j\}$ for $j=1,2,\ldots, n$ and arrange them in the following way:
\begin{align}
    \lambda = (1^{r_1(\lambda)}2^{r_2(\lambda)}\ldots n^{r_n(\lambda)}).\nonumber
\end{align}
Note this implies that the length of the partition $\lambda$ is the sum $\sum_{j=1}^nr_j(\lambda) = l$ and the size of $\lambda$ is the sum $\sum_{j=1}^n jr_j(\lambda) = n$. A graphical representation of such $\lambda$ is called the \textit{Young diagram}. In this paper, we use the English notation which depicts $\lambda$ as the left-aligned row of $\lambda_1$ boxes followed by a row of $\lambda_2$ boxes, and so on until the $l$-th row of $\lambda_l$ boxes. For example, a partition of $7$ with parts $4$, $2$ and $1$ is $(\lambda_1,\lambda_2,\lambda_3) = (4,2,1)$ or $(1^{r_1(\lambda)}2^{r_2(\lambda)}\ldots 7^{r_7(\lambda)}) = (1^12^13^04^15^06^07^0)$ and can be depicted as the following Young diagram:
\begin{align*}
    \yng(4,2,1)
\end{align*}
A Young diagram $\lambda$ of size $n$ and length $l$ covers another Young diagram $\mu$ of size $n-1$ if $\lambda = (\lambda_1,\lambda_2,\ldots,\lambda_l)$ can be obtained by incrementing one of the elements in $\mu = (\mu_1,\mu_2,\ldots,\mu_{l})$ by one in case $\mu$ has a length $l$ or by appending 1 to the end of $\mu = (\mu_1,\mu_2,\ldots,\mu_{l-1})$ if the length of $\mu$ is $l-1$. For example, $\lambda = (4,2,1)$ covers $\mu=(3,2,1)$, $\mu=(4,1,1)$ and $\mu =(4,2)$. Graphically, this amounts to attaching one box to the right-hand side of one of the rows in the Young diagram or attaching one box at the bottom of the leftmost column. We write such $\lambda$ covering $\mu$ as $\lambda\searrow\mu$ or $\mu\nearrow\lambda$.
Finally, the integer partition of $0$ is a special case with the only notation being $\emptyset$.
\subsection{Branching graphs and harmonic functions}\label{sec:BranchingGraph_HarmonicFunction}
A \textit{branching graph} $\mathbb{G} =\coprod_{n=0}^\infty \mathbb{G}_n$ is a disjoint union of partially ordered sets with levels $n\in\mathbb{N}_0$ (see \citealp{stanley2011enumerative} for the definition). The set of elements $\mathbb{G}_n$ in level $n$ is finite and the unique minimal element $\emptyset \in\mathbb{G}_0$ exists while no maximal elements exist in $\mathbb{G}$. We use the notation $|\lambda|$ for the level of the vertex $\lambda \in \mathbb{G}_{|\lambda|}$ (the overlap in notation with Section \ref{sec:partitions_youngdiagram} is intentional and it will be clear in later sections). Furthermore, edges of such $\mathbb{G}$ are oriented and exist only between a pair of vertices $(\mu,\lambda)$ belonging to adjacent levels $\mathbb{G}_{|\mu|}$ and $\mathbb{G}_{|\lambda|}$ where $|\lambda| = |\mu|+1$. The existence of such an edge between $\mu$ and $\lambda$ is denoted as $\mu\nearrow\lambda$ or $\lambda\searrow \mu$ (again, the overlap in notation with \ref{sec:partitions_youngdiagram} is intentional). Finally, we restrict our setting to the connected $\mathbb{G}$ where for any vertex $\mu$ there exists at least one vertex $\lambda\searrow\mu$ and for any $\lambda\neq \emptyset$ there exists at least one $\mu\nearrow\lambda$. For the purpose of reading this paper, a branching graph can be understood as a path space of a nonstationary Markov chain with $\mathbb{G}_n$ being the set of admissible states at time $n$ and edges between the vertices in $\mathbb{G}_n$ and $\mathbb{G}_{n+1}$ as possible transitions of states from time $n$ to $n+1$. 

\begin{figure}[H]
    \centering
    \includegraphics[scale = 0.25]{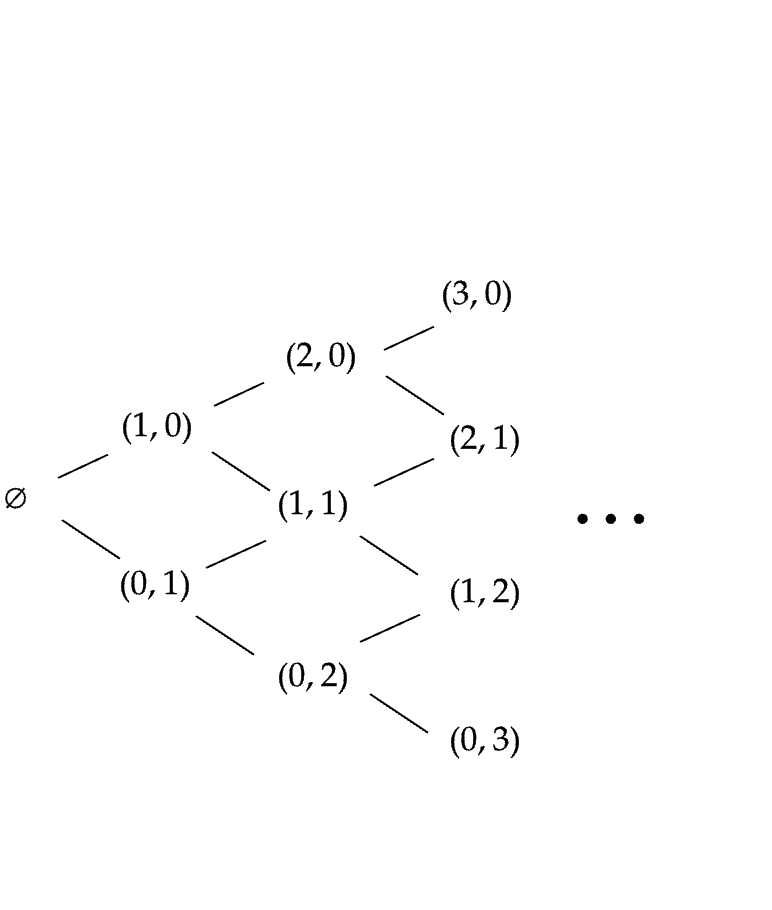}
    \caption{Pascal Triangle}
    \label{fig:PascalTriangle}
\end{figure}

A classical example of a branching graph \citep{kerov1994boundary} is the Pascal triangle viewed as an oriented graph with the $n$-th level consisting of a set:
\begin{align}
    \mathbb{PT}_n = \{(h,t)\in\mathbb{N}^2_{0}:h+t = n\}.\nonumber
\end{align}
Edges of this graph are $(h,t)\nearrow (h+1,t)$ and $(h,t)\nearrow (h,t+1)$ for all $(h,t)$. See 
Figure \ref{fig:PascalTriangle} for a graphical depiction of the Pascal triangle up to the third level. 
Each edge $\mu\nearrow\lambda$ in a branching graph is endowed with a multiplicity function $\varkappa$. It assigns a strictly positive real number $\varkappa(\mu,\lambda)$ (in our setting, it will always be a natural number) between an edge $\mu\nearrow\lambda$. Combined with the aforementioned $\mathbb{G}$, a branching graph is fully determined by the pair $(\mathbb{G},\varkappa)$. A branching graph is called \textit{simple} when $\varkappa$ is 1 for all $(\mu,\lambda)$. The aforementioned Pascal triangle is an example of a simple branching graph. Now, consider an oriented path $\tau = (\lambda{(1)},\lambda{(2)},\ldots,\lambda{(n)})$ of length $n$ on a branching graph starting from a vertex $\lambda(1)$ going through vertices $\lambda(2),\ldots,\lambda(n-1)$ stopping at vertex $\lambda(n)$.  The \textit{weight} of such path $\tau$ is computed multiplicatively $\omega(\tau) = \varkappa(\lambda{(1)},\lambda{(2)})\varkappa(\lambda{(2)},\lambda{(3)})\cdots\varkappa(\lambda{(n-1)},\lambda{(n)})$. The sum of all weights between the vertices $\lambda{(1)}$ and $\lambda{(n)}$ is denoted as 
$$
\dim(\lambda{(1)},\lambda{(n)}) = \sum_{\substack{\tau : \\\text{s.t. }\lambda{(1)}\nearrow \cdots \nearrow \lambda{(n)}}} \omega(\tau)
$$ 
and in particular $\dim(\emptyset,\lambda)$ is called the (combinatorial) \textit{dimension} function which we denote as $\dim(\lambda):= \dim(\emptyset,\lambda)$. The dimension function follows a forward recursion of form
\begin{align}
    \dim(\lambda) = \sum_{\mu:\mu\nearrow \lambda}\varkappa(\mu,\lambda)\dim(\mu).\label{eq:dimrec}
\end{align}
For the Pascal triangle, we have $\dim((a,b),(h,t)) = \binom{h+t-a-b}{t-b}$ and in particular $\dim((h,t)) = \binom{h+t}{t}$. Hence, the forward recursion in \eqref{eq:dimrec} for this graph results in a well known recursion of binomial coefficients $\dim((h,t)) = \dim((h-1,t)) + \dim((h,t-1))$. 

A real-valued function $\varphi$ defined on the set of vertices of a branching graph is called \textit{harmonic} if it satisfies the following backward recursion reciprocal to \eqref{eq:dimrec}:
\begin{align}
    \varphi(\lambda) = \sum_{\nu:\nu\searrow\lambda}\varkappa(\lambda,\nu)\varphi(\nu)\label{eq:harmonicrec}
\end{align}
with a normalization $\varphi(\emptyset) = 1$. We denote the set of all normalized nonnegative harmonic functions on $\mathbb{G}$ as $\mathcal{H}^+_1(\mathbb{G})$. Based on a general result in potential theory \citep{doob1958discrete}, $\varphi\in\mathcal{H}^+_1(\mathbb{G})$ admits an integral representation 
\begin{align}
    \varphi(\lambda) = \int_{\Omega(\mathbb{G})}K(\lambda,\omega)P(d\omega),\qquad \forall\lambda\in\mathbb{G}\label{eq:harmonicint}
\end{align}
on the set of extreme points $\Omega(\mathbb{G})$ in $\mathcal{H}^+_1(\mathbb{G})$ with a probability measure $P$ with a function $K:\mathbb{G}\times \Omega(\mathbb{G})\to \mathbb{R}$. Since the set $\Omega(\mathbb{G})$ is of type $G_\delta$ which is Borel measurable, the probability measure $P$ on $\Omega(\mathbb{G})$ is unique, which makes this representation unique. By choosing a delta measure $P_\omega$ on $\omega\in\Omega(\mathbb{G})$ it follows that $\varphi_\omega(\lambda) = K(\lambda,\omega), \forall \lambda\in\mathbb{G}$. Therefore, the Equation \eqref{eq:harmonicint} states that all harmonic functions are convex combinations of harmonic functions $\varphi_\omega$ on delta measures $P_\omega$ which are called extreme harmonic functions. In all branching graphs we consider in this work, extreme harmonic functions $\varphi_\omega$ can be obtained as the following limit:
\begin{align}
    \varphi_\omega(\lambda) = K(\lambda,\omega) = \lim_{n\to\infty}\frac{\dim({\lambda,\nu(n)})}{\dim(\nu(n))}\label{eq:ergodic}
\end{align}
for a sequence of vertices $\{\nu(n)\in\mathbb{G}_n\}_{n=0}^\infty$ on $\mathbb{G}$.

Harmonic functions $\varphi$ on a branching graph $(\mathbb{G},\varkappa)$ corresponds to a probability function on a sequence of exchangeable random variables whose history at time $n$ corresponds to a vertex in $\mathbb{G}_n$. In other words, all paths  $\tau$ on $\mathbb{G}$ that stops at $\lambda\in\mathbb{G}_n$ represent realizations of random variables of size $n$ equivalent under the exchangeability assumption. For example, paths on the Pascal triangle can be treated as a sequence of binary random variables. Hence, each vertex at $n$-th level respresent the number of heads (or tails) after the $n$-th trial regardless of the order in which heads (and tails) appeared. Using Equation \eqref{eq:ergodic}, the extreme harmonic function for the Pascal triangle is obtained as the following limit for $(a,b)\in\mathbb{PT}_m$ and $(h,t)\in\mathbb{PT}_n$ for $m\leq n$, the latter we treat as a function of $n$ and write instead $(h(n),t(n))$:
\begin{align}
    \lim_{n\to\infty}\frac{\dim((a,b),(h(n),t(n)))}{\dim((h(n),t(n)))} =\lim_{n\to\infty}\frac{\binom{h(n)+t(n)-a-b}{h(n)-a}}{\binom{n}{h(n)}} =\lim_{n\to\infty}\frac{h(n)^{a\downarrow}t(n)^{b\downarrow}}{n^{a+b\downarrow}}\label{eq:extremeharmonic}
\end{align}
This limit exists provided that the limit of the leading terms $\big(\frac{h(n)}{n}\big)^a$ and $\big(\frac{t(n)}{n}\big)^b = \big(1-\frac{h(n)}{n}\big)^b$ exists. Hence, we need the limit of $\frac{h(n)}{n}$ to exist, which we denote $p:= \lim_{n\to\infty}\frac{h(n)}{n}\in[0,1]$. Such $p$ is an element of the set of all extreme points $\Omega(\mathbb{G})$. Hence, $\varphi_p$ is an example of an extreme harmonic function on the Pascal triangle $\mathbb{PT}$. When evaluated at $(a,b)\in\mathbb{PT}_m$, it is given as
\begin{align}
    \varphi_p((a,b)) = K((a,b),p) = \lim_{n\to\infty}\frac{\dim((a,b),(h(n),t(n)))}{\dim((h(n),t(n)))} = p^a(1-p)^b.\nonumber
\end{align}
All extreme harmonic functions $\varphi_\omega$ can be obtained by considering all possible paths $(h(n),t(n))$ on the Pascal triangle where the limit \eqref{eq:extremeharmonic} exists. Hence, $\varphi_\omega$ for all $\omega\in[0,1]$ is the set of all extreme harmonic functions. Therefore, we also establish $\Omega(\mathbb{G}) \equiv [0,1]$.  Thus, the resulting integral representation \eqref{eq:harmonicint} of a harmonic function on the Pascal triangle $\mathbb{PT}$ evaluated at  $(a,b)\in\mathbb{PT}_m$ has a form:
\begin{align}
    \varphi((a,b)) = \int_0^1p^a(1-p)^bP(dp)\nonumber
\end{align}
which is de Finetti's representation theorem on exchangeable binary random variables \citep{deFinetti1979-DEFTOP-2}.
\subsection{Kingman graph and construction of its non-extreme harmonic functions}\label{sec:KingmanHarmonic}
\begin{figure}[H]
    \centering
    \includegraphics[width=0.5\linewidth]{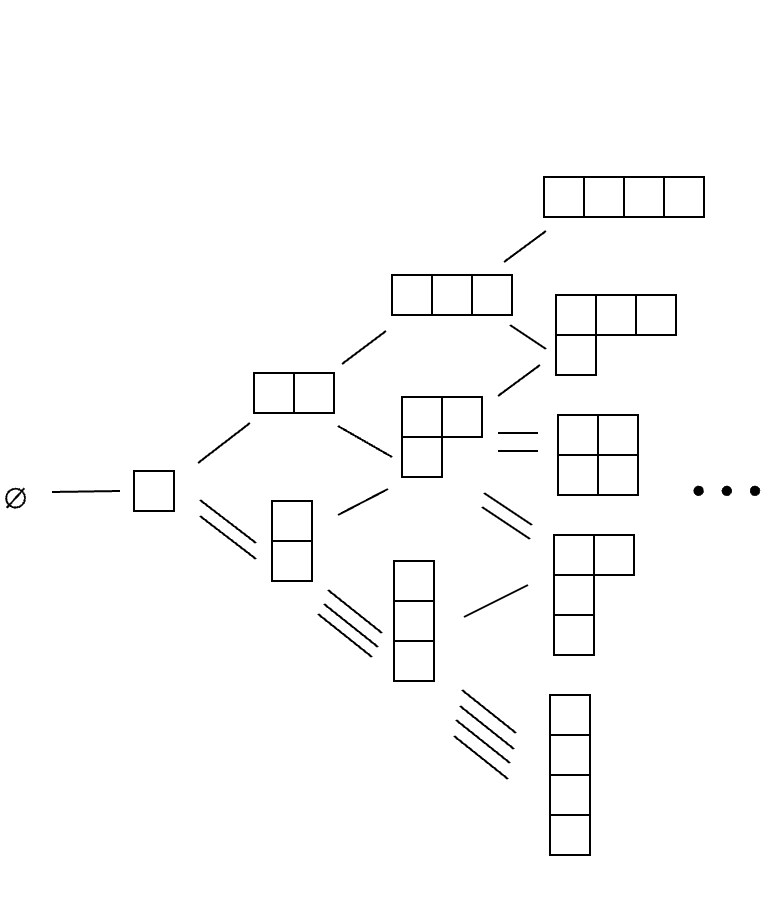}
    \caption{Kingman Graph (the number of edges between vertices corresponds to the multiplicity)}
    \label{fig:KigmanGraph}
\end{figure}

The main topic of this paper, Ewens-Pitman's partition structure is a probability distribution on partitions of the set $\{1,\ldots,n\}$ equivalent under exchangeability. The corresponding branching graph which yields harmonic functions for exchangeable partitions is called the Kigman graph $\mathbb{K}=\coprod_{n=0}^\infty \mathbb{K}_n$ where the vertex set at the $n$-th level is the set of all integer partitions of $n$:
\begin{align}
    \mathbb{K}_n = \{\lambda\in\mathscr{P}_n\}.\nonumber
\end{align}
Therefore, the edges of this branching graph are drawn between a pair of Young diagrams $(\mu,\lambda)$ such that $\lambda$ covers $\mu$ (which explains the overlap in the use of $\nearrow$). The multiplicity function for the Kingman graph is defined as 
\begin{align}
    \varkappa(\mu,\lambda)= r_k(\lambda).\nonumber
\end{align}
That is, the multiplicity is equal to the number of partitions in $\lambda$ which has the exact same number of boxes (including the new box) as the row where the new box $\lambda\setminus\mu$ got attached to $\mu$ to obtain $\lambda$. See Figure \ref{fig:KigmanGraph} for a graphical depiction of this branching graph up to the 4th level.
The term partition structure coined by \cite{kingman1978representation} refers to a sequence $\{M_n\}_{n=1}^\infty$ of coherent probability distributions on set partitions where the distribution $M_{n}$ can be obtained by deleting one of the elements in the distribution of $M_{n+1}$, a property called \textit{consistency} which is crucial for the model to be used in species sampling applications. Consistent $M_n$ follows a recursion of the form \citep{petrov2009two}
\begin{align}
    M_{n}(\lambda) = \sum_{\nu:\nu\searrow\lambda}\frac{\varkappa(\lambda,\nu)\dim(\lambda)}{\dim(\nu)}M_{n+1}(\nu), \qquad |\nu| = |\lambda|+1=n+1\nonumber,
\end{align}
with the normalization $M_0(\emptyset) = 1$. This implies the harmonicity of $M_n(\lambda)/\dim(\lambda)$ by comparing the above with the recursion \eqref{eq:harmonicrec}. Therefore, $M_n(\lambda)$ is given as
\begin{align}
    M_n(\lambda) = \dim(\lambda)\varphi(\lambda)\nonumber
\end{align}
where $|\lambda| = n$, and for the Kingman graph, the dimension function is given as 
\begin{align}
    \dim(\lambda) = \frac{n!}{\lambda_1!\cdots\lambda_{l(\lambda)}!}.\label{eq:dimensionfunction}
\end{align}
As the $\dim(\lambda)$ is determined by the characteristics of the Kingman graph, the law of $M_n(\lambda)$ is fully determined by the harmonic function $\varphi$. Therefore, the harmonic function serves the same role as the exchangeable partition probability function (EPPF,\citealt{pitman1995exchangeable}) that also fully determines the law of $M_n(\lambda)$. In fact, for Ewens-Pitman's partition structure, the harmonic function $\varphi_{\theta,\alpha}(\lambda)$ defined in \eqref{eq:EP_harmonic} differ only with the corresponding EPPF given in \citet[p.~150]{pitman1995exchangeable} by a factor $1/(\prod_{j=1}^nr_j(\lambda)!)$. Therefore, using the analogy with the EPPF, $\varphi(\lambda)$ can be considered the probability on one of the exchangeable paths on the Kingman graph of type $\lambda$ while $M_n(\lambda)$ is the combined probability on all exchangeable paths of type $\lambda$. Hence, we have the normalization
\begin{align}
    \sum_{\lambda\in\mathbb{K}_n}M_n(\lambda) = 1.
\end{align}
From this bijection, one can see that the representation of $M_n$ translates to the representation of $\varphi$ times the multiplicity. We take this viewpoint in dealing with Ewens-Pitman's partition structure which is one of the most commonly used probability distributions on partitions obtained as an induced random partition of the Pitman-Yor process.

When constructing estimators from Ewens-Pitman's partition structure, we need harmonic functions $\varphi$ that are non-extreme as extreme $\varphi$ relates to the asymptotic shape of the exchangeable partitions (see \cite{vershik1981asymptotic} for details, note that they are dealing with the Young graph which is Kingman graph minus the multiplicity). Of course, the integral representation given in \eqref{eq:harmonicint} dictates that non-extreme $\varphi$ can also be represented as a convex combination of extreme harmonic functions. However, rather than relying on this analytical formulation of $\varphi$, we instead utilize the combinatorial and algebraic constructions of nonextreme harmonic function on the Kingman graph given in \cite{borodin1999harmonic} called the interpolation polynomial approach so as to utilize the modern machinery in combinatorics introduced later in this paper.

Let
\begin{align}
    m_\lambda(x_1,x_2,\ldots)= \sum_{(\alpha_1,\alpha_2,\ldots)\sim\lambda}\prod_i x_i^{\alpha_i}\label{def:monomsym}
\end{align}
be the monomial symmetric function indexed by a partition $\lambda$ where the sum is over all sequences of positive integers $(\alpha_1,\alpha_2,\ldots)$ whose elements are rearrangement of nonzero parts of the integer partition $\lambda$ which we denote $(\alpha_1,\alpha_2,\ldots)\sim\lambda$. In \cite{kerov1989combinatorial}, it was pointed out that $m_\lambda$ follows a recursion of the form
\begin{align}
    m_\lambda m_{(1)}  = \sum_{\nu:\nu\searrow\lambda}\varkappa(\lambda,\nu)m_\nu.\nonumber
\end{align}
which suggests the connection with the Kingman graph. Note that $(1)$ is the partition of $1$. It is well known that $\{m_\lambda\}_{\lambda\in\mathbb{K}}$ forms a basis on the algebra of symmetric functions. Another basis of the algebra of symmetric functions related to monomial symmetric functions is factorial monomial symmetric functions $\{m^*_\lambda\}_{\lambda\in\mathbb{K}}$ where each element $m^*_\lambda$ is $m_\lambda$ in \eqref{def:monomsym} with powers $x^{\alpha_i}$ replaced by the descending factorial $x^{\alpha_i\downarrow}$. For example, for an integer partition $\lambda = (2,1)$ with three indeterminates, it is given as
\begin{align}
    m^*_{(2,1)} (x_1,x_2,x_3) &= x_1^{2\downarrow}x_2 + x_1x_2^{2\downarrow} + x_1^{2\downarrow}x_3+x_1x_3^{2\downarrow} + x_2^{2\downarrow}x_3+x_2x_3^{2\downarrow}\nonumber\\
    & = x_1(x_1-1)x_2 + x_1x_2(x_2-1) + x_1(x_1-1)x_3+x_1x_3(x_3-1) + x_2(x_2-1)x_3+x_2x_3(x_3-1)\nonumber
\end{align}
Such $m^*_\lambda$ is the only element on the algebra of symmetric functions that has the highest term equal to the monomial symmetric function $m_\lambda$ and in addition, satisfies 
$
m_\lambda(\mu_1,\mu_2,\ldots) = 0
$ for any $\lambda \neq \mu$ where $|\mu| \leq |\lambda|$. This characteristic is called the \textit{interpolation property} \citep{borodin1999harmonic} of $m^*_\lambda$. Similar to the monomial symmetric function $m_\lambda$, the factorial monomial symmetric function $m^*_\lambda$ also follows a recursion of the form:
\begin{align}
m^*_\lambda m^*_{(1)} =  nm^*_\lambda + \sum_{\nu:\nu\searrow\lambda}\varkappa(\lambda,\nu)m^*_\nu,\qquad |\lambda| = n.\label{eq:factorialmonomrec}
\end{align}
From this recursion \eqref{eq:factorialmonomrec}, it was shown in  \cite{borodin1999harmonic} that $\{m^*_\lambda\}_{\lambda\in\mathbb{K}}$ with the interpolation property can be used to construct non-extreme harmonic functions $\varphi$ on the Kingman graph using the following procedure:
\begin{proposition}\label{prop:harmoniccontruction} \citep{borodin1999harmonic} 
Let $\pi$ be a linear multiplicative functional that maps elements of the algebra of symmetric functions to the real line. Furthermore, let 
\begin{align}
	s = \pi(m^*_{(1)}), \qquad t = -s = -\pi(m^*_{(1)})
\end{align}
such that $s\neq 0,1,2,\ldots$,  then the function
\begin{align}
	\varphi(\lambda) = \frac{\pi(m^*_\lambda)}{s(s-1)\ldots(s-(n-1))} = \frac{(-1)^n\pi(m^*_\lambda)}{t(t+1)\ldots(t+n-1)}, \qquad |\lambda| = n,\label{eq:harmonicinterpolation}
\end{align}
is harmonic in $\mathbb{K}$.
\end{proposition}
The proof is simply applying $\pi$ on both sides of the recursion \eqref{eq:factorialmonomrec}. Then from the multiplicativity and linearity of $\pi$, we obtain the following:
$$
\pi(m^*_\lambda)(s-n) = \sum_{\nu:\nu\searrow\lambda}\varkappa(\lambda,\nu)\pi(m^*_\nu).
$$
Finally, by dividing both sides by $s(s-1)\ldots(s-n)$ we obtain $\varphi$ of the form \eqref{eq:harmonicinterpolation} which follows the recursion \eqref{eq:harmonicrec}, thus harmonic.

 A special case of the general method given in Proposition \ref{prop:harmoniccontruction}  with $\theta = t$ and 
\begin{align}
	\pi_{\theta,\alpha}(m^*_\lambda) = (-1)^{|\lambda|}\frac{\theta^{l(\lambda)\uparrow\alpha}}{\prod_{i=1}^{|\lambda|}r_i(\lambda)!}\prod_{j=1}^{l(\lambda)}(1-\alpha)^{\lambda_j-1\uparrow}.\label{eq:numSignedEwensPitman}
\end{align}
results in a harmonic function $\varphi_{\theta,\alpha}$ in Equation \eqref{eq:EP_harmonic}
\begin{align}
    \varphi_{\theta,\alpha}(\lambda) = \frac{(-1)^n\pi_{\theta,\alpha}(m^*_\lambda)}{\theta^{n\uparrow}}, \qquad |\lambda| = n.\label{eq:ewensharmonic}
\end{align}
This harmonic function $\varphi_{\theta,\alpha}(\lambda)$ times the multiplicity $\dim(\lambda)$ is Ewens-Pitman's partition structure $M_n(\lambda)$. To see that such $\pi_{\theta,\alpha}$ is indeed linear and mulitiplicative with respect to $m^*_\lambda$, we first note that $\pi_{-k\alpha,\alpha}(m^*_\lambda)$ for $\theta = -k\alpha, k =1,2,\ldots$ coincides with the following:
\begin{align}
	m^*_\lambda(\underbrace{\alpha,\ldots,\alpha}_{k}) = \frac{k^{l(\lambda)\downarrow}}{\prod_{i=1}^{|\lambda|}r_i(\lambda)!}\prod_{j=1}^{l(\lambda)}\alpha^{\lambda_j\downarrow}.\label{eq:defmstar}
\end{align}
Therefore, it is multiplicative for $\theta = -k\alpha, k = 1,2,\ldots$. Now, we can view $\pi_{\theta,\alpha}$ in  \eqref{eq:numSignedEwensPitman} as a polynomial of $\theta$ with fixed $\lambda$ and $\alpha$. This polynomial is denoted as $f_\lambda(\theta;\alpha):= \pi_{\theta,\alpha}(m^*_\lambda)$ and it is multiplicative for all values of $\theta$ and given $\lambda$ and $\alpha$. This means that one can construct $f_\lambda(\theta;\alpha)$ and thereby Ewens-Pitman's partition structure by interpolating this polynomial $f_\lambda(\theta;\alpha)$ through nodes $\theta = -l(\lambda)\alpha, -(l(\lambda)+1)\alpha, \ldots$ and the corresponding values in $f_\lambda(\theta;\alpha)$ which are 
$$
f_\lambda(-l(\lambda)\alpha;\alpha) = m^*_\lambda(\underbrace{\alpha,\ldots,\alpha}_{l(\lambda)}), f_\lambda(-(l(\lambda)+1)\alpha;\alpha) = m^*_\lambda(\underbrace{\alpha,\ldots,\alpha}_{l(\lambda)+1}),\ldots.
$$
Note that we start from $\theta = -l(\lambda)\alpha$ because the term $k^{l(\lambda)\downarrow}$ will be zero in \eqref{eq:defmstar} for $\theta = -k\alpha$ with $k$ less than $l(\lambda)$.
\subsection{Newton polynomial interpolation of $f_\lambda(\theta;\alpha)$}\label{sec:NewtonPoly}

For a partition $\lambda\in\mathbb{K}_n$ of size $|\lambda| = n$ with length $l(\lambda) = l$, the polynomial $f_\lambda(\theta;\alpha)$ has a degree $l$. Therefore, points 
$$
 (-l\alpha, m^*_\lambda(\underbrace{\alpha,\ldots,\alpha}_{l})), (-(l+1)\alpha, m^*_\lambda(\underbrace{\alpha,\ldots,\alpha}_{l+1})),\ldots, (-(2l-1)\alpha, m^*_\lambda(\underbrace{\alpha,\ldots,\alpha}_{2l-1}))
$$
are used to interpolate $f_\lambda(\theta;\alpha)$. The Newton polynomial interpolation of $f_\lambda(\theta;\alpha)$ through these points will have the following form:
\begin{align}
	f_\lambda(\theta;\alpha) = &f[-l\alpha] + f[-l\alpha,-(l+1)\alpha](\theta+l\alpha) + f[-l\alpha,-(l+1)\alpha,-(l+2)\alpha](\theta+l\alpha)(\theta+(l+1)\alpha)+\cdots + \nonumber\\
	&f[-l\alpha,-(l+1)\alpha,\ldots,-2l\alpha](\theta+l\alpha)(\theta+(l+1)\alpha)\cdots(\theta+(2l-1)\alpha)\label{eq:NewtonForm}
\end{align}
where $f[-l\alpha,\ldots,-(l+i)\alpha], i = 0,1,\ldots,l$ are divided differences obtained through the following procedure. First, an initialization with $f[-(l+i)\alpha] = f_\lambda(-(l+i)\alpha;\alpha) = m^*_\lambda(\underbrace{\alpha,\ldots,\alpha}_{l+i})$ for $i=0,\ldots,l$ will take place. Then, $f[-l\alpha,\ldots,-(l+i)\alpha]$ can be obtained recursively as follows:
\begin{align}
	 f[-l\alpha,\ldots,-(l+i)\alpha] = \frac{f[-(l+1)\alpha,\ldots,-(l+i)\alpha]-f[-l\alpha,\ldots,-(l+i-1)\alpha]}{-(l+i)\alpha+l(\alpha)}.\label{eq:iteration}
\end{align}

For example, to obtain
\begin{align}
    \varphi_{\theta,\alpha}(\lambda = (2,2,1)) = \frac{\theta^{3\uparrow \alpha}}{2!\theta^{5\uparrow }}(1-\alpha)^2\label{eq:ewensexample}
\end{align}
we first initialize 
$$
f[-3\alpha] = m^*_{(2,2,1)}(\underbrace{\alpha,\ldots,\alpha}_{3}),
f[-4\alpha] = m^*_{(2,2,1)}(\underbrace{\alpha,\ldots,\alpha}_{4}),
f[-5\alpha] = m^*_{(2,2,1)}(\underbrace{\alpha,\ldots,\alpha}_{5}),
f[-6\alpha] = m^*_{(2,2,1)}(\underbrace{\alpha,\ldots,\alpha}_{6})
$$
and obtain
$$
f[-3\alpha,-4\alpha] = \frac{m^*_{(2,2,1)}(\underbrace{\alpha,\ldots,\alpha}_{4})-m^*_{(2,2,1)}(\underbrace{\alpha,\ldots,\alpha}_{3})}{-\alpha}.
$$
Similarly, $f[-4\alpha,-5\alpha]$ and $f[-5\alpha,-6\alpha]$ can be computed in the same way. Iterating this procedure as in \eqref{eq:iteration} for $f[-3\alpha,-4\alpha,-5\alpha]$, $f[-4\alpha,-5\alpha,-6\alpha]$, and finally $f[-3\alpha-4\alpha,-5\alpha,-6\alpha]$, we obtain our special case of Equation \eqref{eq:NewtonForm} as follows:
\begin{align}
    f_{(2,2,1)}(\theta;\alpha) =& f[-3\alpha] + f[-3\alpha,-4\alpha](\theta+3\alpha) + f[-3\alpha,-4\alpha,-5\alpha](\theta+3\alpha)(\theta+4\alpha)\nonumber\\&+f[-3\alpha-4\alpha,-5\alpha,-6\alpha](\theta+3\alpha)(\theta+4\alpha)(\theta+5\alpha)\nonumber\\
    =& \frac{(-1)^5}{2!}\theta^{3\uparrow\alpha}(1-\alpha)^2\label{example:Newton}
\end{align}
Plugging in the value of $f_{(2,2,1)}(\theta;\alpha) = \pi_{\theta,\alpha}(m^*_{(2,2,1)})$ given as \eqref{example:Newton} to the formula of $\varphi_{\theta,\alpha}$ in Equation \eqref{eq:ewensharmonic} gives the value in \eqref{eq:ewensexample}. This calculation is cumbersome by hand but can be easily done by any computer algebra software.

Importantly, Equation \eqref{eq:NewtonForm} makes it clear that Ewens-Pitman's partition structure can be constructed using a linear multiplicative map on factorial monomial symmetric functions $m^*_\lambda$ as operations carried out in finite differencing are all linear with respect to them. Constructing Ewens-Pitman's partition structure in this way enables the use of modern combinatorial machinery which will be introduced in the next section. These modern tools can be used to obtain various estimators in a principled manner without relying on ad-hoc combinatorial techniques.

\section{Riordan Arrays and Modern Umbral Interpolation via Sheffer Sequences}\label{sec:RiordanArrayShefferSeq}
The construction of $f_\lambda(\theta;\alpha)$ via Newton polynomial interpolation shown in Section \ref{sec:NewtonPoly} is a special case of an expansion of polynomials using Sheffer sequences called (modern) umbral interpolation. In this section, we first introduce Sheffer sequences in their original form proposed in \cite{sheffer1939} and then show the characterization used in \cite{roman1984}. The latter under the operation called ``umbral composition'' forms a group called Sheffer group which is isomorphic to a group of infinite lower-triangular matrices called Riordan arrays \citep{shapiro1991riordan}. We show that this correspondence can be used to compute certain marginals of Ewens-Pitman's partition structure via Riordan array sums utilizing the fundamental theorem of Riordan arrays (FTRA). Finally, by phrasing the Newton polynomial interpolation of $f_\lambda(\theta;\alpha)$ as an umbral interpolation problem, we give a new representation of $f_\lambda(\theta;\alpha)$ and thereby that of Ewens-Pitman's partition structure.
\subsection{Expansion via Sheffer A-type zero polynomial sequence}\label{sec:AtypeZeroExpansion}
Let $\mathcal{P}_l$ be the space of polynomials with degrees less than or equal to $l$ for $l\in\mathbb{N}$. We expand the polynomial $f_\lambda(\theta;\alpha)\in\mathcal{P}_{l(\lambda)}$ for $\lambda\in\mathbb{K}_n$ using a sequence of polynomials $\{P_i(\theta;\alpha,\lambda)\}_{i=0}^\infty$ (where the polynomial $P_i(\theta;\alpha,\lambda)$ is of degree $i$) of form
\begin{align}
    P_i(\theta;\alpha,\lambda):= \frac{(\theta+l(\lambda)\alpha)^{i\uparrow\alpha}}{i!} = (\theta+l(\lambda)\alpha)(\theta+(l(\lambda)+1)\alpha)\cdots(\theta+(l(\lambda)+i-1)\alpha)/(i!)\nonumber
\end{align}
where $\theta$ is the variable while $\alpha$ and $\lambda$ are parameters. This sequence of polynomials $\{P_i(\theta;\alpha,\lambda)\}_{i=0}^\infty$ is an example of a Sheffer A-type zero polynomial sequence \citep{sheffer1939} 
characterized by the generating function equation 
\begin{align}
    A(t)e^{\theta H(t)} = \sum_{i=0}^\infty P_i(\theta;\alpha,\lambda) t^i\label{eq:ShefferSeq}
\end{align}
with
\begin{align}
    A(t) = \sum_{i=0}^\infty a_it^i\in\mathcal{F}_0
    ,\qquad H(t) = \sum_{i=0}^\infty h_it^i\in\mathcal{F}_1\nonumber
\end{align}
where $\mathcal{F}_0,\mathcal{F}_1$ are the subset of the ring of formal power series $\mathcal{F}$ for the real field
\begin{align}
    \mathcal{F} = \Big\{\sum_{i=0}^\infty f_it^i | f_i\in\mathbb{R} \Big\}.\nonumber
\end{align}
and are defined as 
\begin{align}
    \mathcal{F}_0 = \Big\{f\in \mathcal{F}| f_0 \neq  0\Big\},
    \qquad \mathcal{F}_1  = \Big\{f\in\mathcal{F}|f_0 = 0,f_1\neq 0\} = t\mathcal{F}_0.\nonumber
\end{align}
By specializing $A(t)$ and $H(t)$ as follows
\begin{align}
    A(t) = (1-\alpha t)^{-l(\lambda)},
    \qquad H(t) = \frac{-\log(1-\alpha t)}{\alpha}\nonumber
\end{align}
we obtain the form of $P_i(\theta;\alpha,\lambda):= \frac{(\theta+l(\lambda)\alpha)^{i\uparrow\alpha}}{i!}$ used in our setting. Sheffer discovered that for such $P_i(\theta;\alpha,\lambda)$, there exists a unique operator $J$ whose coefficients can be stored in a formal power series of form
\begin{align}
    J(t) = \sum_{i=0}^\infty f_i t^i \in\mathcal{F}_1.\nonumber
\end{align}
which in operator form $J(D)$ (where $D$ is the formal derivative operator with respect to the variable $\theta$) acts on $P_i(\theta;\alpha,\lambda)$ by
\begin{align}
    J(D)P_i(\theta;\alpha,\lambda) = P_{i-1}(\theta;\alpha,\lambda)\label{eq:ShefferLoweringOp}.
\end{align} 
This operator, in the formal power series form $J(t)$ is a compositional inverse of $H(t)$. That is, $J(H(t)) = H(J(t)) = t$. Hence, in our setting, it specializes to
\begin{align}
    J(t) = \frac{1-\exp(-\alpha t)}{\alpha}.\nonumber
\end{align}
Therefore, the operator $J(D)$ can be written as follows:
\begin{align}
    J(D) = \frac{1-\exp(-\alpha D)}{\alpha} = \frac{1-E^{-\alpha}}{\alpha} = \frac{\nabla_\alpha}{\alpha}\nonumber
\end{align}
where $E$ is the shift operator which acts on $f(\theta)$ by $Ef(\theta) = f(\theta+1)$, hence $E^{-\alpha}f(\theta) = f(\theta-\alpha)$ while $\nabla_\alpha$ is the backward difference operator with step size $\alpha$ which acts on $f(\theta)$ by $\nabla_\alpha f(\theta) = f(\theta)-f(\theta-\alpha)$. Furthermore, we used the relation $\exp(D) = E$ between the difference operator $D$ and the shift operator $E$. Readers unfamiliar with these results may refer to \cite{jordan1965calculus}. With this, one can easily check that the specialization of the Equation \eqref{eq:ShefferLoweringOp} is
\begin{align}
    \frac{\nabla_\alpha}{\alpha}\frac{(\theta+l(\lambda)\alpha)^{i\uparrow\alpha}}{i!} = \frac{(\theta+l(\lambda)\alpha)^{i-1\uparrow\alpha}}{(i-1)!}.\nonumber
\end{align}
This result implies the following identity:
\begin{align}
    \Bigg(\frac{\nabla_\alpha}{\alpha}\Bigg)^j \frac{(\theta+l(\lambda)\alpha)^{i\uparrow\alpha}}{i!}\Bigr|_{\theta = -l(\lambda)\alpha} = \delta_{i,j}\label{eq:ShefferDelta}
\end{align}
where $\delta_{i,j} = 1$ if $i=j$ and 0 otherwise. Hence, $f_\lambda(\theta;\alpha)$ of degree $l(\lambda)$ can be expanded over the basis $(\theta+l(\lambda)\alpha)^{i\uparrow\alpha}, i = 0,1,\ldots, l(\lambda),$ as follows:
\begin{align}
    f_\lambda(\theta;\alpha) = \sum_{i=0}^{l(\lambda)}b_i (\theta+l(\lambda)\alpha)^{i\uparrow\alpha},\nonumber
\end{align}
with coefficients \begin{align}
    b_i = \frac{1}{i!\alpha^i}\nabla^i_\alpha f_\lambda(\theta;\alpha)\Bigr|_{\theta = -l(\lambda)\alpha}, \qquad i = 0,1,\ldots,l(\lambda).\nonumber
\end{align}
One can verify that such $b_i$ is indeed the divided difference used in the Newton polynomial form of $f_\lambda(\theta;\alpha)$ in Equation \eqref{eq:NewtonForm} by first noting that:
\begin{align}
    \frac{1}{i!\alpha^i}\nabla^i_\alpha f_\lambda(\theta;\alpha)=f[\theta,\theta-\alpha,\ldots,\theta-i\alpha]\nonumber
\end{align}
which is the well-known relation between divided difference with equal step size and the finite difference. Therefore, $b_i$ is given as
\begin{align}
    b_i =\frac{1}{i!\alpha^i}\nabla^i_\alpha f_\lambda(\theta;\alpha)\Bigr|_{\theta = -l(\lambda)\alpha} = f[-l(\lambda)\alpha,-(l(\lambda)+1)\alpha,\ldots,-(l(\lambda)+i)\alpha].\nonumber
\end{align}

With this, we established that the Newton polynomial form of $f_\lambda(\theta;\alpha)$ given in Equation \eqref{eq:NewtonForm} is an expansion of $f_\lambda(\theta;\alpha)$ in terms of Sheffer A-type zero polynomial sequence. In the subsequent subsections, we introduce a more modern characterization of the Sheffer sequence used in \cite{roman1984} which enables us to utilize the Riordan array, a recent tool in combinatorics, to compute some summary statistics and estimators derived from Ewens-Pitman's partition structure.
\subsection{The Riordan array and its generalizations}
Consider a pair of formal power series $(d(t), h(t))$:
\begin{align}
    d(t) &= \sum_{i=0}^\infty d_it^i \in\mathcal{F}_0\nonumber\\
    h(t) &= \sum_{i=1}^\infty h_it^i\in\mathcal{F}_1\nonumber
\end{align}
Let $(d_{n,k})_{n,k\in\mathbb{N}_0}$ be an infinite lower-triangular matrix over $\mathbb{R}$. If there exists a pair $(d(t),h(t))$ such that the $(n,k)$-th element of the matrix is defined as:
\begin{align}
    d_{n,k} = [t^n]d(t)h(t)^k,\nonumber
\end{align}
where $[t^n]$ is an operator that takes the coefficient of the term $t^n$, then, such a matrix is called a Riordan array and we denote it as $\mathcal{R}(d(t),h(t))$. If $\mathcal{R}(d(t),h(t))$ and $\mathcal{R}(g(t),f(t))$ are Riordan arrays, the multiplication $*$ is defined as:
\begin{align}
    \mathcal{R}(d(t),h(t))*\mathcal{R}(g(t),f(t)):= \mathcal{R}(d(t)g(h(t)),f(h(t))\nonumber
\end{align}
which is another Riordan array. Hence, the identity matrix in Riordan array form is $\mathcal{R}(1,t)$ while the inverse matrix of $\mathcal{R}(d(t),h(t))$ is given as:
\begin{align}
    \mathcal{R}(d(t),h(t))^{-1}:= \mathcal{R}\Bigg(\frac{1}{d(\bar{h}(t))},\bar{h}(t)\Bigg)\nonumber
\end{align}
where $\bar{h}(t)$ is the compositional inverse of $h(t)$, that is $h(\bar{h}(t))=\bar{h}(h(t))=t$. Therefore, the set of Riordan arrays forms a group called the Riordan group under multiplication.

While Riordan arrays have various useful properties, we primarily utilize the following result called the fundamental theorem of Riordan arrays (FTRA):
\begin{theorem}\citep{shapiro1991riordan}: Every combinatorial sum representable as a linear combination of the elements in the $n$-th row $d_{n,k},k=0,\ldots$ of Riordan array $\mathcal{R}(d(t),h(t))$ can be written down as follows:
\begin{align}
    \sum_{k=0}^\infty d_{n,k}c_k = \sum_{k=0}^nd_{n,k}c_k = [t^n]d(t)c(h(t))\nonumber
\end{align}
where $c(t) = \sum_{i=0}^\infty c_it^i$. 
\end{theorem}
The type of Riordan array used primarily in this paper is called the \textit{Exponential Riordan array}, which is a generalization of the Riordan array. It is again defined using a pair of power series $(d(t),h(t))$, but this time exponential power series:
\begin{align}
     d(t) &= \sum_{i=0}^\infty d_i\frac{t^i}{i!}, \qquad d_0\neq0, \qquad d_i\in\mathbb{R},\nonumber \\
    h(t) &= \sum_{i=1}^\infty h_i\frac{t^i}{i!},\qquad h_1\neq 0,\qquad h_i\in\mathbb{R}.\nonumber   
\end{align}
From here on, we denote the exponential Riordan array defined with respect to exponential power series $d(t)$ and $h(t)$ as $\mathcal{R}_e[d(t),h(t)]$. The $(n,k)$-th element of $\mathcal{R}_e[d(t),h(t)]$ is defined as 
\begin{align}
    d_{n,k}=\Big[\frac{t^n}{n!}\Big]d(t)\frac{h(t)^k}{k!}.\nonumber
\end{align} 
Note that the extraction of the coefficient for the term $\frac{t^n}{n!}$ on an exponential power series has a nice interpretation as an operation of taking the $n$-th derivative of the series with respect to $t$ followed by setting $t=0$. For example, $d_2 = \big[\frac{t^2}{2!}\big]d(t) = \frac{d}{dt}^2d(t)\bigr|_{t=0}$. 

The FTRA with respect to $\mathcal{R}_e[d(t),h(t)]$ and an exponential series $c(t) =\sum_{i=0}^\infty c_i\frac{t^i}{i!}$ is defined as
\begin{align}
    \sum_{k=0}^\infty d_{n,k}c_k = \sum_{k=0}^nd_{n,k}c_k = \Big[\frac{t^n}{n!}\Big]d(t)c(h(t))\label{eq:eFTRA}
\end{align}
\subsection{The Riordan group and the Sheffer group}\label{sec:RGroupSSeq}
The expansion of the polynomial $f_\lambda(\theta;\alpha)$ via the Sheffer A-type zero sequence \citep{sheffer1939} introduced in Section \ref{sec:AtypeZeroExpansion} is encompassed in the modern theory of umbral interpolation \citep{costabile2025umbral} initiated by the work \cite{roman1978umbral} and the book of \cite{roman1984} as the main reference for further developments.
Consider a sequence of polynomials $\{s_i(\theta)\}_{i=0}^\infty$ (where $s_i(\theta)$ is of degree $i$) for a pair of characteristic exponential power series $(d(t),h(t))$ (overlap in notation with the previous section is intentional) with real-valued coefficients provided that $d_0\neq0$, $h_0=0$ and $ h_1\neq 0$. We define such $s_i(\theta),i = 0,1,\ldots$ as the coefficient of the following exponential generating function \cite[p.~18]{roman1984}:
\begin{align}
    \frac{1}{d(\bar{h}(t))}e^{\theta\bar{h}(t)} = \sum_{i=0}^\infty s_i(\theta)\frac{t^i}{i!}.\label{eq:Shefferseq2}
\end{align}
where $\bar{h}(t)$ is the compositional inverse of $h(t)$. This sequence $\{s_i(\theta)\}_{i=0}^\infty$ is called the Sheffer sequence \citep{roman1984}. The Sheffer sequence and the Sheffer A-type zero polynomial sequence defined in Equation \eqref{eq:ShefferSeq} coincide. Specifically, $\{s_i(\theta)\}_{i=0}^\infty$ is a Sheffer sequence in Roman's notation\footnote{There is another notation without the term $i!$ used in works such as \cite{he2007sheffer}. Throughout this work, we stick to Roman's notation.} if and only if $\{\frac{1}{i!}s_i(\theta)\}_{i=0}^\infty$ is a Sheffer A-type zero sequence \citep{roman1978umbral,costabile2022towards}. From here on, we deal exclusively with the Sheffer sequence $\{s_i(\theta)\}_{i=0}^\infty$ and translate results involving Sheffer A-type zero sequence established in Section \ref{sec:AtypeZeroExpansion} whenever necessary.

A set of Sheffer sequences $\{s_i(\theta)\}_{i=0}^\infty$ forms a group under the operation \textit{umbral composition} which we denote as ``$\circ$''. An umbral composition between a sequence $\{s_i(\theta)\}_{i=0}^\infty$ Sheffer for $(d(t),h(t))$ and  a sequence $\{r_i(\theta)\}_{i=0}^\infty$ Sheffer for $(g(t),f(t))$ is defined as another sequence $\{r_i(\theta)\circ s_i(\theta)\}$ Sheffer for $(d(t)g(h(t),f(h(t)))
$. One can easily check that the Sheffer sequence for $(d(t)=1,h(t)=t)$ serves as the identity and the inverse is given by the conjugate Sheffer sequence $\{\hat{s}_i(\theta)\}_{i=0}^\infty$ with the exponential generating function\footnote{In newer works, Equation \eqref{eq:conjugateSheffer} rather than \eqref{eq:Shefferseq2} is used as the definition of the Sheffer sequence for $(g(t),f(t))$, see for example \cite{gould2013characterization} and  \cite{costabile2014algebraicSheffer}. In this work, as in \cite{shapiro2022riordan}, we stick to the definition given by \cite{roman1984} and make necessary changes to results stated in other papers that follow different conventions.}:
\begin{align}
    d(t)e^{\theta h(t)} = \sum_{i=0}^\infty \hat{s}_i(\theta)\frac{t^i}{i!}.\label{eq:conjugateSheffer}
\end{align}
Viewing these results in terms of its characteristic exponential power series, it appears to be identical to the multiplication rule of matrices, identity matrix and inverse matrix of the exponential Riordan array in the previous section. Indeed, the isomorphism between the Riordan group and the Sheffer group was established by \cite{he2007sheffer}. While we leave the details of this isomorphism to \cite{he2007sheffer}, the following result will be needed in later sections. From the Equation \eqref{eq:conjugateSheffer}, we have
\begin{align}
    \hat{s}_i(\theta) = \Big[\frac{t^i}{i!}\Big]d(t)e^{\theta h(t)}
    =\sum_{j=0}^\infty\Big[\frac{t^i}{i!}\Big] d(t)(h(t))^j\frac{\theta^j}{j!}
    =\sum_{j=0}^i d_{i,j}\theta^j
    =\sum_{j=0}^i d_{i,j}\Big[\frac{t^j}{j!}\Big]e^{\theta t}\label{eq:conjugateSheffer2}
\end{align}
where $d_{i,j}$ is the $(i,j)$-th cell of the exponential Riordan array $\mathcal{R}_e[d(t),h(t)]$. These equations can be recognized as a special case of the FTRA involving exponential power series given in \eqref{eq:eFTRA}. That is, the $i$-th conjugate Sheffer series for $(d(t),h(t))$ can be obtained as an $i$-th coefficient of the power series one obtains as a result of carrying out FTRA with respect to the $i$-th row of the exponential Riordan array $\mathcal{R}_e[d(t),h(t)]$ via exponential power series $c(t) = e^{\theta t}$.
\subsection{Modern umbral interpolation and the FTRA}\label{sec:UmbralInterpolationFTRA}
Now, consider a sequence $\{s_i(\theta)\}_{i=0}^\infty$ Sheffer for $(d(t),h(t))$ with the generating function given as Equation \eqref{eq:Shefferseq2}. Then, there exists a linear operator $Q$ which acts on $f(\theta)\in\mathcal{P}_l, l \in\mathbb{N}_0$ of form:
\begin{align}
    Q = \sum_{i=0}^\infty h_i\frac{D^i}{i!},\qquad h_0=0, \qquad h_1\neq0,\nonumber
\end{align}
where $h_i, i =0,1,\ldots$ are the coefficients of the $h(t)$ in the generating function of the conjugate Sheffer sequence $\{\hat{s}_i(\theta)\}_{i=0}^\infty$ and $D$ is the formal derivative operator with respect to $\theta$. Such operator is called the \textit{$\delta$-operator} and has the following degree reducing property:
\begin{align}
    Qs_l(\theta) = ls_{l-1}(\theta), \qquad l=1,2,\ldots.\nonumber
\end{align}
One can easily check this result by setting $Q=J(D)$ in Equation \eqref{eq:ShefferLoweringOp} and by in noting the aforementioned correspondence between the Sheffer sequence $\{s_i(\theta)\}_{i=0}^\infty$ and the Sheffer A-type zero sequence $\{\frac{1}{i!}s_i(\theta)\}_{i=0}^\infty$. Furthermore, let $L$ be a linear functional on $\mathcal{P}_l, l\in\mathbb{N}_0$ such that $L(1) \neq 0$. Then, there exists such a functional $L$ which in conjunction with $Q$ satisfies the following
\begin{align}
    L(Q^js_l(\theta)) = j!\binom{l}{j}\delta_{j,l},\qquad j = 1,2,\ldots,l,\nonumber
\end{align}
which enables $\{s_0(\theta),s_1(\theta)\ldots,s_l(\theta)\}$ to be basis on $\mathcal{P}_l$ \citep{costabile2025umbral}. This is called $\{s_i(\theta)\}_{i=0}^\infty$ being \textit{umbral basis} for a pair $(L,Q)$. From this result, umbral interpolation can be carried out using the following theorem and corollary which first appeared in \cite{costabile2014algebraic} that generalizes the expansion in Section \ref{sec:AtypeZeroExpansion}:
\begin{theorem}\citep{costabile2025umbral}. Let $\{s_i(\theta)\}_{i=0}^\infty$ be the umbral basis for  a given $(L,Q)$. Then, the polynomial 
\begin{align}
    q_l(\theta) = \sum_{j=0}^l\omega_js_j(\theta)\nonumber
\end{align}
is the unique polynomial of degree less than or equal to $l$ that satisfies the umbral interpolation problem. That is, for the polynomial $q_l\in\mathcal{P}_l$, there exists
\begin{align}
    L(Q^jq_l) = j!\omega_j,\nonumber
\end{align}
where $\omega_j\in\mathbb{R}$ for $j=0,1,\ldots,l$.
\end{theorem}
\begin{corollary}\label{thr:RepPoly} (Representation of Polynomials, \citealp{costabile2025umbral}) Each $q_l\in\mathcal{P}_l$ can be written as
\begin{align}
    q_l(\theta) = \sum_{j=0}^l L(Q^jq_l)\frac{s_j(\theta)}{j!}.\nonumber
\end{align}
\end{corollary}
We specialize the Corollary \ref{thr:RepPoly} to our setting where $f_\lambda(\theta;\alpha)$ will be treated as a polynomial of degree $l(\lambda)$ for any given $\lambda\in\mathbb{K}_n, n = 1,2,\ldots$ and $\alpha$. This polynomial $f_\lambda(\theta;\alpha)$ is expanded over $s_i(\theta) = (\theta+l(\lambda)\alpha)^{i\uparrow\alpha}, i = 0,1,\ldots,l(\lambda)$. Then, we have $Q= \nabla_\alpha/\alpha$ and $L$ is an evaluation functional at $\theta = -l(\lambda)\alpha$ as covered in Section  \ref{sec:AtypeZeroExpansion}. Thus, Corollary \ref{thr:RepPoly} specializes to the following representation:
\begin{align}
    f_\lambda(\theta;\alpha) &= \sum_{j=0}^{l(\lambda)}\frac{L(\nabla_\alpha^j f_\lambda(\theta;\alpha))}{\alpha^jj!}(\theta+l(\lambda)\alpha)^{j\uparrow\alpha}\label{eq:fRepresentation1_2}.
\end{align}
Now, it is no coincidence that the first term in the sum \eqref{eq:fRepresentation1_2} can be written as follows
\begin{align}
    \frac{L(\nabla_\alpha^j f_\lambda(\theta;\alpha))}{\alpha^jj!} = \frac{(d(D)h(D)^jf_\lambda(\theta;\alpha))\bigr|_{\theta=0}}{j!}\nonumber
\end{align}
where $d(D) = E^{-l\alpha}$ and $h(D) = \frac{1-E^{-\alpha}}{\alpha}$ are operator form of exponential power series $d(t) = \exp(-l\alpha t)$ and $h(t) = \frac{1-\exp(-\alpha t)}{\alpha}$ that generates the conjugate Sheffer sequence $\{\hat{s}_i(\theta)\}_{i=0}^\infty\}$ using Equations \eqref{eq:conjugateSheffer} and \eqref{eq:conjugateSheffer2}. This is because the pair $(d(t),h(t))$ and $(L,Q)$ define each other (for detailed reasons, we refer to \citealt{roman1984}). Then, as $d_0\neq 0$ in $d(t)$ and $h_0=0, h_1\neq0 $ in $h(t)$, the operator $d(D)h(D)^j$ reduces the degree of $f_\lambda(\theta;\alpha)$ minimally by $j$ resulting in $d(D)h(D)^jf_\lambda(\theta;\alpha) = 0$ for $j=l+1,l+2,\ldots$. Thus, Equation \eqref{eq:fRepresentation1_2} can be rewritten as follows with summations now extended to infinity:
\begin{align}
    f_\lambda(\theta;\alpha) &= \sum_{j=0}^\infty (d(D)h(D)^jf_\lambda(\theta;\alpha))\bigr|_{\theta=0}\frac{(\theta+l(\lambda)\alpha)^{j\uparrow\alpha}}{j!}.\label{eq:fexpansion}
\end{align}
We use this expanded form of $f_\lambda(\theta;\alpha)$ to show that the marginalization of $f_\lambda(\theta;\alpha)$ with respect to all partitions $\lambda\in\mathbb{K}_n$ of length $l(\lambda) = l$, times the signed multiplicity term $(-1)^n\dim(\lambda)$ can be written as a Riordan array sum (times the constant outside the sum) which we can use FTRA to compute.

The aforementioned marginalization of \eqref{eq:fexpansion} proceeds as follows. First, we exchange the order of the summation:
\begin{align}   \sum_{\substack{\lambda\in\mathbb{K}_n\\l(\lambda)=l}}(-1)^n\dim(\lambda)f_\lambda(\theta;\alpha) = \sum_{j=0}^\infty\frac{(\theta+l\alpha)^{j\uparrow\alpha}}{\alpha^j}\sum_{\substack{\lambda\in\mathbb{K}_n\\l(\lambda)=l}}\frac{(-1)^n\dim(\lambda)(E^{-l\alpha}\nabla_\alpha^jf_\lambda(\theta;\alpha))\bigr|_{\theta=0}}{j!}.\label{eq:EPmarginal} 
\end{align}
Then, observe that the term $(\theta+l\alpha)^{j\uparrow\alpha}/\alpha^j$ is the $j$-th coefficient of the exponential generating function $(1-t)^{-\theta/\alpha-l}$. Therefore, if the inner sum with respect to all partitions $\lambda\in\mathbb{K}_n$ of length $l(\lambda) = l$ has an exponential Riordan array representation, we can treat \eqref{eq:EPmarginal} as a generalization of \eqref{eq:conjugateSheffer2} and find its generating function using the FTRA. To verify this, let's expand the inner sum as follows:
\begin{align}
\sum_{\substack{\lambda\in\mathbb{K}_n\\l(\lambda)=l}}\frac{(-1)^n\dim(\lambda)(E^{-l\alpha}\nabla_\alpha^jf_\lambda(\theta;\alpha))\bigr|_{\theta=0}}{j!}=\sum_{\substack{\lambda\in\mathbb{K}_n\\l(\lambda)=l}}\frac{(-1)^n}{j!}\frac{n!}{\lambda_1!\lambda_2!\cdots\lambda_l!}\sum_{i=0}^j(-1)^i\binom{j}{i}f_\lambda(-(l+i)\alpha;\alpha)\label{eq:shefferD_1}
\end{align}
where we used the definition of the dimension function on $\lambda\in\mathbb{K}_n$ in \eqref{eq:dimensionfunction}, commutation relation between $E$ and $\nabla$ and the binomial theorem when expanding with respect to the operator $\nabla^j_\alpha = (1-E^{-\alpha})^j$. Then, as established in Section \ref{sec:KingmanHarmonic}
\begin{align}
    f_\lambda(-(l+i)\alpha;\alpha) = \pi_{-(l+i)\alpha,\alpha}(m^*_\lambda) = m^*_\lambda(\underbrace{\alpha,\ldots,\alpha}_{l+i}) =\frac{(l+i)^{l\downarrow}}{\prod_{i=1}^{n}r_i(\lambda)!}\prod_{j=1}^{l}\alpha^{\lambda_j\downarrow}\nonumber
\end{align}
and the summation over partitions $\lambda\in\mathbb{K}_n$ with $l(\lambda) = l$ can be converted to summation over compositions $c_1+\cdots + c_l = n$ times the multiplicity $({\prod_{i=1}^{n}r_i(\lambda)!})/l!$, Equation \eqref{eq:shefferD_1} becomes
\begin{align} \frac{(-\alpha)^l l!}{j!}\sum_{i=0}^j(-1)^i\binom{j}{i}\frac{(l+i)^{i\downarrow}}{i!} \sum_{c_1+\cdots+c_l = n}\frac{n!}{l!}\prod_{j=1}^l\frac{(1-\alpha)^{c_j-1\uparrow}}{c_j!}.\label{eq:marginalization1}
\end{align}
Here, we can recognize the inner sum in \eqref{eq:marginalization1} as the generalized Stirling number $S_{n,l}(-1,-\alpha,0)$ which is a special case of $S_{n,l}(a,b,c)$ given in \cite{hsu1998unified}. To see this, here is the explicit form of $S_{n,l}(a,b,0)$ which equals a particular specialization of the exponential partial Bell polynomial$B_{n,k}$
\begin{align}
    S_{n,l}(a,b,0) = B_{n,l}(\omega_1,\ldots,\omega_{n+l-1}) = \sum_{c_1+\cdots+c_l =n}\frac{n!}{l!}\prod_{j=1}^l\frac{\omega_{c_j}}{c_j!}\label{eq:GSN_Bell}
\end{align}
where $\omega_i = (b-a)^{i-1\downarrow a}$. We use the notation $S_{n,l}^{a,b,c}:=S_{n,k}(a,b,c)$ from here on. Taking $S_{n,l}^{-1,-\alpha,0}$ outside the outer sum of \eqref{eq:marginalization1}, we carry out FTRA on the remaining terms inside the sum and after some simplifications, we obtain the following exponential Riordan array representation of \eqref{eq:marginalization1} and thereby \eqref{eq:shefferD_1} (details can be found in the Appendix \ref{app:appA}):
\begin{align}
    S_{n,l}^{-1,-\alpha,0}\frac{(-\alpha)^l l!}{j!}\sum_{i=0}^j(-1)^i\binom{j}{i}\frac{(l+i)^{i\downarrow}}{i!} 
     &=S_{n,l}^{-1,-\alpha,0} (-\alpha)^l\Big[\frac{t^l}{l!}\Big]\Big(\frac{-1}{t-1}\Big)\frac{\Big(\frac{t}{t-1}\Big)^j}{j!}\label{eq:FTRAinner_5}.
\end{align} 
Thus, Equation \eqref{eq:EPmarginal} can be written as a form of FTRA \eqref{eq:eFTRA} as follows:
\begin{align}
\sum_{\substack{\lambda\in\mathbb{K}_n\\l(\lambda)=l}}(-1)^n\dim(\lambda)f_\lambda(\theta;\alpha) &= S^{-1,-\alpha,0}_{n,l}(-\alpha)^l\sum_{j=0}^\infty \Big[\frac{t^j}{j!}\Big](1-t)^{-\theta/\alpha-l}\Big[\frac{t^l}{l!}\Big]\Big(\frac{-1}{t-1}\Big)\frac{\Big(\frac{t}{t-1}\Big)^j}{j!}\nonumber\\
&=S^{-1,-\alpha,0}_{n,l}(-\alpha)^l\Big[\frac{t^l}{l!}\Big]\Big(\frac{-1}{t-1}\Big)\Big(1-\frac{t}{t-1}\Big)^{-\theta/\alpha-l}\nonumber\\
&=S^{-1,-\alpha,0}_{n,l}\theta^{l\uparrow\alpha}.\label{eq:FTRAmarginal}
\end{align}
In Section \ref{Sec:Estimators}, we use this result to obtain various summary statistics and estimators derived from Ewens-Pitman's partition structures in a unified manner.
\subsection{A new representation of Ewens-Pitman's partition structure}\label{sec:altrep}
From the specialization of Collorary \ref{thr:RepPoly} in our setting given in \eqref{eq:fRepresentation1_2}, a more fundamental form of $f_\lambda(\theta;\alpha)$ can be obtained by swapping the order of summations as follows:
\begin{align}
    f_\lambda(\theta;\alpha) & = \sum_{j=0}^{l(\lambda)}\frac{\sum_{i=0}^j(-1)^i\binom{j}{i}f_\lambda(\theta-i\alpha;\alpha)\bigr|_{\theta = -l(\lambda)\alpha}}{\alpha^jj!}(\theta+l(\lambda)\alpha)^{j\uparrow\alpha}\nonumber\\
    &= \sum_{j=0}^{l(\lambda)}f_\lambda(\theta-j\alpha;\alpha)\bigr|_{\theta = -l(\lambda)\alpha}\sum_{i=0}^{l(\lambda)-j}(-1)^j\binom{i+j}{j}\frac{(\theta+l(\lambda)\alpha)^{i+j\uparrow\alpha}}{\alpha^{i+j}(i+j)!}\label{eq:swappedrep}.
\end{align}
A similar method of swapping the sums for the operator $Q=\Delta_\alpha$ case was introduced in \cite{costabile2013delta}.
After further simplifications of \eqref{eq:swappedrep}, one can prove the following representation of $f_\lambda(\theta;\alpha)$ (proof in Appendix \ref{appendix:B}):
\begin{align}
f_\lambda(\theta;\alpha)=\sum_{j=0}^{l(\lambda)}m^*_\lambda(\underbrace{\alpha,\ldots,\alpha}_{l(\lambda)+j})(-1)^j\frac{(\theta+l(\lambda)\alpha)^{l(\lambda)+1\uparrow\alpha}}{(\theta+(l(\lambda)+j)\alpha)\alpha^{l(\lambda)}(l(\lambda)-j)!j!}\label{eq:fAltRepresentation}.
\end{align}
Therefore, by using the definition of $\varphi_{\theta,\alpha}(\lambda)$ given in \eqref{eq:ewensharmonic} and noting that $f_\lambda(\theta;\alpha):=\pi_{\theta,\alpha}(m^*_\lambda)$, the new representation of Ewens-Pitman's partition structure is given as follows:
\begin{align}
    M_n(\lambda) = \dim(\lambda)\varphi_{\theta,\alpha}(\lambda) = \frac{n!(-1)^n}{\prod_{i=1}^{l(\lambda)}\lambda_i!\theta^{n\uparrow}}\sum_{i=0}^{l(\lambda)}m^*_\lambda(\underbrace{\alpha,\ldots,\alpha}_{l(\lambda)+i})(-1)^i\frac{(\theta+l(\lambda)\alpha)^{l(\lambda)+1\uparrow\alpha}}{(\theta+(l(\lambda)+i)\alpha)\alpha^{l(\lambda)}(l(\lambda)-i)!i!}
\end{align}
for $\lambda\in\mathbb{K}_n$. This representation, although implied by \cite{borodin1999harmonic} to the best of our knowledge is the first explicit form of Ewens-Pitman's partition structure as a linear combination of factorial monomial symmetric functions $m^*_\lambda$. Importantly, this representation separates the term $m^*_\lambda$ which enumerates the weights given on the partition sizes $(\lambda_1,\lambda_2,\ldots)$ using only the parameter $\alpha$ and the term proportional to the partition length $l(\lambda)$ which not only depends on $\alpha$ but also $\theta$. This aligns with the known property of the Ewens-Pitman's partition structure that the partition length $l(\lambda)$ is the sufficient statistics of $\theta$ while conditional on $l(\lambda)$ the partition sizes $(\lambda_1,\lambda_2,\ldots)$ is the sufficient statistics of $\alpha$ mentioned in works such as \cite{pitman2006combinatorial}, \cite{10.1214/15-STS529} and \cite{mano2018partitions}.
\section{Summary Statistics and Estimators of Ewens-Pitman’s Partition Structure Using the FTRA}\label{Sec:Estimators}
In Section \ref{sec:UmbralInterpolationFTRA}, we showed that the term 
\begin{align}
    \sum_{\substack{\lambda\in\mathbb{K}_n\\l(\lambda)=l}}(-1)^n\dim(\lambda)f_\lambda(\theta;\alpha),\nonumber
\end{align}
which is the marginal of the unnormalized Ewens-Pitman partition function $$
M_n(\lambda) = \dim(\lambda)\varphi_{\theta,\alpha}(\lambda) = \dim(\lambda)\frac{(-1)^nf_\lambda(\theta;\alpha)}{\theta^{n\uparrow}}
$$ can be written as a form of FTRA times the generalized Stirling number $S_{n,l}^{-1,\alpha,0}$. Furthermore, the generalized Stirling number $S_{n,l}^{a,b,c}$ has a Riordan array representation as follows:
\begin{align}
    S_{n,l}^{a,b,c}=\Big[\frac{t^n}{n!}\Big](1+\alpha t)^{\frac{c}{a}}\frac{\Big(\frac{(1+at)^{\frac{b}{a}}-1}{b}\Big)^l}{l!}.\label{eq:GSNRiordan}
\end{align}
That is, $S_{n,l}^{a,b,c}$ can be obtained from the $(n,l)$-th cell of the exponential Riordan array $\mathcal{R}_e[(1+\alpha t)^{\frac{c}{a}},\frac{(1+at)^{\frac{b}{a}}-1}{b}]$. Given this, we show an FTRA-based approach to obtain summary statistics and estimators derived from marginals of the Ewens-Pitman's partition structure. All marginals in this section are proportional to the form
\begin{align}
\sum_{\lambda\in\mathbb{K}_n} g(\lambda)(-1)^n\dim(\lambda)f_\lambda(\theta;\alpha)\nonumber
\end{align}
for some integer partition valued function $g(\lambda)$ which may include an indicator function to restrict the range of summation. The list of possible functional forms of $g(\lambda)$ given in this section is not meant to be exhaustive.

To motivate the simplicity of this approach, we first consider a simple case where $g(\lambda) = 1$ for all $\lambda\in\mathbb{K}_n$. We know that this quantity is the normalization constant $\theta^{n\uparrow}$ of the Ewens-Pitman's partition structure as 
\begin{align}
    \sum_{\lambda\in\mathbb{K}_n}M_n(\lambda) = \sum_{\lambda\in\mathbb{K}_n}\dim(\lambda)\varphi_{\theta,\alpha}(\lambda)  = \sum_{\lambda\in\mathbb{K}_n}\dim(\lambda)\frac{(-1)^nf_\lambda(\theta;\alpha)}{\theta^{n\uparrow}} = 1\nonumber
\end{align}
the probability must sum up to one. We can easily verify this quantity through an elementary application of FTRA as follows:
\begin{align}
    \sum_{\lambda\in\mathbb{K}_n}(-1)^n\dim(\lambda)f_\lambda(\theta;\alpha) &= \sum_{l=0}^n\sum_{\substack{\lambda\in\mathbb{K}_n\\l(\lambda)=l}}(-1)^n\dim(\lambda)f_\lambda(\theta;\alpha)\label{eq:FTRAnormalization1}\\
    &=\sum_{l=0}^n S_{n,l}^{-1,-\alpha,0} \theta^{l\uparrow\alpha}\label{eq:FTRAnormalization2}\\
    &= \sum_{l=0}^n\Big[\frac{t^n}{n!}\Big]\frac{\Big(\frac{1-(1-t)^\alpha}{\alpha}\Big)^l}{l!}\Big[\frac{t^l}{l!}\Big](1-\alpha t)^{-\frac{\theta}{\alpha}}\label{eq:FTRAnormalization3}\\
    &=\Big[\frac{t^n}{n!}\Big](1-t)^{-\theta} = \theta^{n\uparrow}\label{eq:FTRAnormalization4}
\end{align}
where from \eqref{eq:FTRAnormalization1} to \eqref{eq:FTRAnormalization2}, we used the central result in Section \ref{sec:UmbralInterpolationFTRA} given in Equation \eqref{eq:FTRAmarginal} while from \eqref{eq:FTRAnormalization2} to \eqref{eq:FTRAnormalization3} we used the Riordan array representation of the $S_{n,l}^{a,b,c}$ in Equation \eqref{eq:GSNRiordan} and the exponential generating function $(1-\alpha t)^{-\frac{\theta}{\alpha}}$ of $\theta^{l\uparrow\alpha}$. The final result in \eqref{eq:FTRAnormalization4} then follows from the FTRA.
\subsection{Factorial moments of partition lenghts}\label{sec:MomentPartitionLength}
We start with the factorial moments of order $k$ ($k = 1,2,\ldots,l(\lambda)-1$) for the partition length $l(\lambda)$ when $\lambda\in\mathbb{K}_n$. For Ewens-Pitman's partition structure, \cite{yamato2000moments} derived these summary statistics via recursion. In our setting, it amounts to specifying $g(\lambda) = l(\lambda)^{k\downarrow}$. The target quantity can then be written as follows:
\begin{align}
     \mathbb{E}_n[l(\lambda)^{k\downarrow}] &= \frac{1}{\theta^{n\uparrow}}\sum_{\lambda\in\mathbb{K}_n}l(\lambda)^{k\downarrow}(-1)^n\dim(\lambda)f_\lambda(\theta;\alpha)\nonumber\\
     &= \frac{1}{\theta^{n\uparrow}}\sum_{l=0}^nl^{k\downarrow}\sum_{\substack{\lambda\in\mathbb{K}_n\\l(\lambda)=l}}(-1)^n\dim(\lambda)f_\lambda(\theta;\alpha)\nonumber\\
     &=\frac{1}{\theta^{n\uparrow}}\sum_{l=0}^nl^{k\downarrow}\theta^{l\uparrow\alpha}S_{n,l}^{-1,-\alpha,0}\nonumber
\end{align}
Then, we utilize the following derivative rule for the generating function involving formal derivative operator $D_t$ with respect to $t$
\begin{align}
    \sum_{i=0}^\infty i^{k\downarrow} f_i\frac{t^i}{i!} = \sum_{i=0}^\infty (tD_t)^kf_i\frac{t^i}{i!}\label{eq:trick}.
\end{align}
This rule combined with the Riordan array representation of $S_{n,l}^{-1,-\alpha,0}$ in \eqref{eq:GSNRiordan} and the exponential generating function $(1-\alpha t)^{-\frac{\theta}{\alpha}}$ of $\theta^{l\uparrow\alpha}$ results in the following:
\begin{align}
    \mathbb{E}_n[l(\lambda)^{k\downarrow}]
    &=\frac{1}{\theta^{n\uparrow}}\sum_{l=0}^n\Big[\frac{t^l}{l!}\Big](tD_t)^k(1-\alpha t)^{-\frac{\theta}{\alpha}}\Big[\frac{t^n}{n!}\Big]\frac{\Big(\frac{1-(1-t)^\alpha}{\alpha}\Big)^l}{l!}\nonumber.
\end{align}
Taking the formal derivative $D_t$ $k$ times and rearranging some terms followed by the FTRA gives
\begin{align}
\mathbb{E}_n[l(\lambda)^{k\downarrow}]=\frac{\theta^{k\uparrow\alpha}}{\theta^{n\uparrow}}\Big[\frac{t^n}{n!}\Big]\Big(\frac{1-(1-t)^\alpha}{\alpha}\Big)^k(1-t)^{-\theta-k\alpha}.\nonumber
\end{align}
Now, observe that the right-hand side terms of the coefficient extraction operator are the generating function of the generalized Stirling number in Riordan array representation \eqref{eq:GSNRiordan} minus the factor $\frac{1}{k!}$. With this, we can obtain the target quantity shown in \cite{yamato2000moments} as follows:
\begin{align}
\mathbb{E}_n[l(\lambda)^{k\downarrow}]&=\frac{k!\theta^{k\uparrow\alpha}}{\theta^{n\uparrow}}S_{n,k}^{-1,-\alpha,\theta+k\alpha}\nonumber\\
    & = \Big(\frac{\theta}{\alpha}\Big)^{k\uparrow}\sum_{j=0}^k(-1)^{k-j}\binom{k}{j}\frac{(\theta+j\alpha)^{n\uparrow\alpha}}{\theta^{n\uparrow}}\nonumber
\end{align}
where the last equation is obtained using the expansion of the generalized Stirling number given in \cite{he2013expression}. 
\subsection{Joint factorial moments of partition sizes}\label{sec:JointFacPartitionSizes}
The joint factorial moments of order $(k_1,k_2\ldots,k_n)$ for partition sizes in frequency notation $(1^{r_1(\lambda)}2^{r_2(\lambda)}\ldots n^{r_n(\lambda)})$ (provided that $n>\sum_{j=1}^n jk_j$) for Ewens-Pitman's partition structure are given in \cite{yamato2000moments}. For a more general Gibbs-type model, those are given in \cite{cerquetti2013marginals} and \cite{favaro2013conditional}. These statistics in our setting can be obtained by specifying $g(\lambda) = \prod_{j=1}^n r_j(\lambda)^{k_j\downarrow}$. First, the target quantity can be rewritten as follows:
\begin{align}
    \mathbb{E}_n\Big[\prod_{j=1}^n r_j(\lambda)^{k_j\downarrow}\Big] &= \frac{1}{\theta^{n\uparrow}}\sum_{\lambda\in\mathbb{K}_n}\prod_{j=1}^n r_j(\lambda)^{k_j\downarrow}(-1)^n\dim(\lambda)f_\lambda(\theta;\alpha)\nonumber\\
    &= \frac{1}{\theta^{n\uparrow}}\sum_{l=0}^n\prod_{j=1}^n r_j(\lambda)^{k_j\downarrow}\sum_{\substack{\lambda\in\mathbb{K}_n\\l(\lambda)=l}}(-1)^n\dim(\lambda)f_\lambda(\theta;\alpha)\nonumber\\
    &=\frac{1}{\theta^{n\uparrow}}\sum_{l=0}^n\prod_{j=1}^n r_j(\lambda)^{k_j\downarrow}\theta^{l\uparrow\alpha}S_{n,l}^{-1,-\alpha,0}\label{eq:jointfacmoment1}
\end{align}
Then, upon noting that 
 \begin{align}
     n = \sum_{j=1}^njr_j(\lambda), \qquad l(\lambda) = \sum_{j=1}^nr_j(\lambda)\nonumber
 \end{align}
 we rewrite $S_{n,l}^{-1,-\alpha,0}$ in frequency notation whose definition as an exponential partial Bell polynomial is given in \eqref{eq:GSN_Bell} 
\begin{align}
   S_{n,l}^{-1,-\alpha,0} = \sum_{\substack{\sum_{j=1}^njr_j(\lambda)=n\\
   \sum_{j=1}^nr_j(\lambda) =l}}\prod_{j=1}^n\Big(\frac{(1-\alpha)^{j-1\uparrow}}{j!}\Big)^{r_j(\lambda)}\frac{n!}{r_j(\lambda)!}
\end{align}
where the factor $\frac{l!}{\prod_{j=1}^n r_j(\lambda)}$ is introduced when converting the summation over compositions to partitions, the rest follows a change of variables. Once $S_{n,l}^{-1,-\alpha,0}$ is in frequency notation, the derivative rule \eqref{eq:trick} can be used again to \eqref{eq:jointfacmoment1} which gives the following
\begin{align}
    \mathbb{E}_n\Big[\prod_{j=1}^n r_j(\lambda)^{k_j\downarrow}\Big] = \frac{1}{\theta^{n\uparrow}}\sum_{l = 0}^n \prod_{j=1}^n(\xi_j D_{\xi_j})^{k_j}\theta^{l\uparrow\alpha}\sum_{\substack{\sum_{j=1}^njr_j(\lambda)=n\\
   \sum_{j=1}^nr_j(\lambda) =l}}\prod_{j=1}^n\xi_j^{r_j(\lambda)}\frac{n!}{r_j(\lambda)!}\label{eq:jointfacmoment2}
\end{align}
with $\xi_j = \frac{(1-\alpha)^{j-1\uparrow}}{j!}$. Now we introduce terms $s=\sum_{j=1}^n j k_j$ and $k = \sum_{j=1}^nk_j$ and simplify above \eqref{eq:jointfacmoment2} as follows:
\begin{align}
    \mathbb{E}_n\Big[\prod_{j=1}^n r_j(\lambda)^{k_j\downarrow}\Big] = \frac{n!\prod_{j=1}^n\xi_j^{k_j}}{(n-s)!\theta^{n\uparrow}}\sum_{l-k = 0}^{n-s} \theta^{l\uparrow\alpha}\sum_{\substack{\sum_{j=1}^njr_j(\lambda)=n-s\\
   \sum_{j=1}^nr_j(\lambda) =l-k}}\prod_{j=1}^n\xi_j^{r_j(\lambda)-k_j}\frac{(n-s)!}{(r_j(\lambda)-k_j)!}\label{eq:jointfacmoment3}
\end{align}
note the multiplication and division by $(n-s)!$. Observe that the inner sum can be recognized as the generalized Stirling number $S_{n-s,l-k}^{-1,-\alpha,0}$. Now all we need is to bring the outer sum of \eqref{eq:jointfacmoment3} in the form of FTRA. This can be achieved by factorizing $\theta^{l\uparrow\alpha}$ into a product of $(\theta)^{k\uparrow\alpha}$ and $(\theta+k\alpha)^{l-k\uparrow\alpha}$ of which latter stays in the sum as $l-k$-th coefficient of the generating function $(1-\alpha t)^{-\frac{\theta}{\alpha}-k}$. Hence, we have the following FTRA form
\begin{align}
    \mathbb{E}_n\Big[\prod_{j=1}^n r_j(\lambda)^{k_j\downarrow}\Big] = \frac{n!\prod_{j=1}^n\xi_j^{k_j}}{(n-s)!\theta^{n\uparrow}}\theta^{k\uparrow\alpha}\sum_{l-k = 0}^{n-s} \Big[\frac{t^{l-k}}{(l-k)!}\Big](1-\alpha t)^{-\frac{\theta}{\alpha}-k}\Big[\frac{t^{n-s}}{(n-s)!}\Big]\frac{\big(\frac{1-(1-t)^\alpha}{\alpha}\big)^{l-k}}{(l-k)!}\nonumber.
\end{align}
Carrying out the FTRA will give the target quantity shown in \cite{yamato2000moments}
\begin{align}
     \mathbb{E}_n\Big[\prod_{j=1}^n r_j(\lambda)^{k_j\downarrow}\Big] = \frac{n!}{(n-s)!}\frac{\theta^{k\uparrow\alpha}(\theta+k\alpha)^{n-s\uparrow}}{\theta^{n\uparrow}}\prod_{j=1}^n\Big(\frac{(1-\alpha)^{j-1\uparrow}}{j!}\Big)^{k_j}.
\end{align}
While the intermediate steps required before the FTRA are rather long, the overall procedure is essentially the same as in Section \ref{sec:MomentPartitionLength}. That is, we construct terms that are constant inside the sum and the remaining terms should either be the coefficient of an exponential power series or have a Riordan array representation. Then, FTRA is applied to obtain the target quantity.
\subsection{Joint conditional factorial moments of partition sizes}
Estimators of Ewens-Pitman's partition structure of size $n+m$ conditional on a partition of size $n$ have an important application in species sampling models in Bayesian nonparametrics (see for example \citealt{lijoi2008bayesian} and \citealt{favaro2013conditional}). Specifically, consider a partition $\lambda\in\mathbb{K}_{n+m}$ which is obtained by adding $m$ extra observations to a partition $\mu= (\mu_1,\ldots,\mu_{l(\mu)})\in\mathbb{K}_n$. In the language of branching graphs, we consider all paths of class $\lambda$ to that goes through $\mu$. Here, we show that one of the estimators derived from $\lambda$ conditional on $\mu$ given in \cite{cerquetti2013marginals} and \cite{favaro2013conditional} can be obtained using the FTRA. 

Let $\lambda\in\mathbb{K}_{n+m}$ be written as a disjoint union of the following frequencies
\begin{align}
(1^{\rold_1(\lambda)}2^{\rold_2(\lambda)}\ldots(n+m)^{\rold_{n+m}(\lambda)})\  \coprod \ (1^{\rnew_1(\lambda\setminus\mu)}2^{\rnew_2(\lambda\setminus\mu)}\ldots m^{\rnew_{m}(\lambda\setminus\mu)})\label{eq:LambdaDisjoint}
\end{align}
where the left is the frequency of counts in parts that belong to $\lambda\in\mathbb{K}_{n+m}$ which also belong to $\mu\in\mathbb{K}_n$ (albeit with potentially different nonzero counts) while the right is the frequency of counts in parts that did not exist in $\mu$ which belong to $\lambda\setminus\mu$. The former can potentially consist of one part with all $n+m$ observations, hence its frequency ranges between $1$ to $n+m$. The latter may consist of one part with all $m$ additional observations, thus ranging between $1$ to $m$.

We are interested in the factorial moments of order $(k_1,k_2,\ldots,k_m)$ on new partition sizes in frequency notation $(1^{\rnew_1(\lambda\setminus\mu)}2^{\rnew_2(\lambda\setminus\mu)}\ldots m^{\rnew_{m}(\lambda\setminus\mu)})$ that belong to $\lambda\in\mathbb{K}_{n+m}$ (provided that $\sum_{j=1}^mj\rnew_j(\lambda) > \sum_{j=1}^m jk_j$) conditional on $\mu = (\mu_1,\ldots,\mu_{l(\mu)})\in\mathbb{K}_n$ for Ewens-Pitman's partition structure. This estimator is defined slightly different from those in \cite{cerquetti2013marginals} and \cite{favaro2013conditional} as those conditioned on partition sizes $(n_1,\ldots,n_{l(\mu)})$ in exchangeable order, i.e., $(n_1,\ldots,n_{l(\mu)})\sim \mu$. However, it will result in the same quantity (as it will result in the appearance of the same ordering factor $\frac{1}{l!}$ in both the numerator and the denominator of the quantity we compute). In our setting where we view this problem as an aggregation of paths on the Kingman graph, conditioning on the vertex $\mu\in\mathbb{K}_n$ makes more sense. 

In this case, the function $g(\lambda)$ is $\prod_{j=1}^m\rnew_j(\lambda\setminus\mu)^{k_j\downarrow}$ times the indicator function that the path goes through $\mu$ when entering the $n$-th level $\mathbb{K}_n$. This simply results in the marginalization over $\lambda\in\mathbb{K}_{n+m}$ to be conditional on $\mu\nearrow\cdots\nearrow\lambda$. Hence, the target quantity can be written as follows:
\begin{align}
    \mathbb{E}_{n+m}\Big[\prod_{j=1}^m\rnew_j(\lambda\setminus\mu)^{k_j\downarrow}\Bigr|\mu\Big] &= \sum_{\substack{\lambda\in\mathbb{K}_{n+m}\\
    \text{s.t. }\mu\nearrow\cdots\nearrow\lambda}}
    \prod_{j=1}^m\rnew_j(\lambda\setminus\mu)^{k_j\downarrow} \frac{M_{n+m}(\lambda)}{M_n(\mu)}\nonumber\\
    &=\frac{\theta^{n\uparrow}}{\theta^{n+m\uparrow}}\sum_{\substack{\lambda\in\mathbb{K}_{n+m}\\
    \text{s.t. }\mu\nearrow\cdots\nearrow\lambda}}
    \prod_{j=1}^m\rnew_j(\lambda\setminus\mu)^{k_j\downarrow} \frac{(-1)^{n+m}\dim(\lambda)f_\lambda(\theta;\alpha)}{(-1)^n\dim(\mu)f_\mu(\theta;\alpha)}\label{eq:cond_moment}
\end{align}
Now, observe that marginalization takes place for all paths that go through $\mu$. For those, $\dim(\lambda)$ factorizes as follows:
\begin{align}
    \dim(\lambda) = \dim(\mu)\dim(\lambda\setminus\mu)\nonumber.
\end{align}
The latter term has a form
\begin{align}
    \dim(\lambda\setminus\mu) = \frac{(m-\rnew)!\rnew!}{q_1!\cdots q_{l(\mu)}!\prod_{j=1}^{\rnew}(j!)^{\rnew_j(\lambda\setminus\mu)}}\nonumber
\end{align}
where $q_1,\ldots,q_{l(\mu)}$ are nonnegative (zero is possible) increments on $(\mu_1,\ldots,\mu_{l(\mu)})$ so that it becomes $(1^{\rold_1(\lambda)}2^{\rold_2(\lambda)}\ldots(n+m)^{\rold_{n+m}(\lambda)})$ in frequency notation while we define $\rnew$ as
$$
\rnew = \sum_{j=1}^mj\rnew_j(\lambda\setminus\mu).
$$ I.e., the number of observations that belong to newly created parts. Furthermore, we set 
$$
l^\text{new} =\sum_{j=1}^m\rnew_j(\lambda\setminus\mu)
$$
as the number of new parts created with $m$ additional samples. After some simplifications and the derivative rule detailed in Appendix \ref{appendix:cond_details}, we obtain the following form of \eqref{eq:cond_moment}.
\begin{align}
    \frac{\theta^{n\uparrow}}{\theta^{n+m\uparrow}}\sum_{l =0}^m \prod_{j=1}^m(\xi_jD_{\xi_j})^{k_j}\frac{\theta^{l+l(\mu)\uparrow\alpha}}{\theta^{l(\mu)\uparrow\alpha}}\sum_{r=0}^m
    \Bigg(\sum_{\substack{
    \rnew = r\\
    l^\text{new} = l
    }}\prod_{j=1}^r\xi_j^{\rnew_j(\lambda\setminus\mu)}\frac{r!}{\rnew_j(\lambda\setminus\mu)!}\Bigg)\Bigg(
    \sum_{\substack{q_1+\cdots+q_{l(\mu)}\\
    =m-r}}
    \binom{m-r}{q_1,\ldots,q_{l(\mu)}}\prod_{j=1}^{l(\mu)}(\mu_j-\alpha)^{q_j\uparrow}
    \Bigg).\label{eq:cond_moment2}
\end{align}
where $\xi_j = \frac{(1-\alpha)^{j-1\uparrow}}{j!}$. Observe that the first term in the big bracket is $S_{r,l}^{-1,-\alpha,0}$, and the second term in the big bracket can be simplified to $(n-l(\mu)\alpha)^{m-r}$ using a multivariate version of Chu–Vandermonde identity \citep[Appendix A.2]{lijoi2008bayesian}. Since both $S_{r,l}^{-1,-\alpha,0}$ and $(n-l(\mu)\alpha)^{m-r}$ can be written as coefficients of a generating function, the inner sum of \eqref{eq:cond_moment2} can be written as follows:
\begin{align}
    \sum_{r=0}^m\Big[\frac{t^r}{r!}\Big]\frac{\big(\frac{1-(1-t)^\alpha}{\alpha}\big)^{l}}{(l)!}\Big[\frac{t^{m-r}}{(m-r)!}\Big](1-t)^{-n+l(\mu)\alpha} 
    &= \Big[\frac{t^m}{m!}\Big](1-t)^{-n+l(\mu)\alpha}\frac{\big(\frac{1-(1-t)^\alpha}{\alpha}\big)^{l}}{(l)!}\label{eq:convolution}\\
    &= S_{m,l}^{-1,-\alpha,n-l(\mu)\alpha}\label{eq:convolution2}
\end{align}
where the convolution rule of generating functions is used to obtain \eqref{eq:convolution}. Then, \eqref{eq:convolution2} follows from the Riordan array form of generalized Stirling number given in Equation \eqref{eq:GSNRiordan}.
Now, we introduce terms $s = \sum_{j=1}^mjk_j$ and $k=\sum_{j=1}^mk_j$ and plug in \eqref{eq:convolution2} to \eqref{eq:cond_moment2} and obtain the following:
\begin{align}
    \mathbb{E}_{n+m}\Big[\prod_{j=1}^m\rnew_j(\lambda\setminus\mu)^{k_j\downarrow}\Bigr|\mu\Big] 
    &=\frac{\theta^{n\uparrow}}{\theta^{n+m\uparrow}}\prod_{j=1}^m\xi_j^{k_j}\sum_{l=0}^m\prod_{j=1}^mD_{\xi_j}^{k_j}\frac{\theta^{l+l(\mu)\uparrow\alpha}}{\theta^{l(\mu)\uparrow\alpha}}S_{m,l}^{-1,-\alpha,n-l(\mu)\alpha}\nonumber\\
    &=\frac{m^{s\downarrow}\theta^{n\uparrow}}{\theta^{n+m\uparrow}}\prod_{j=1}^m\xi_j^{k_j}\sum_{l-k=0}^{m-s}(\theta+l(\mu)\alpha)^{l\uparrow\alpha}S_{m-s,l-k}^{-1,-\alpha,n-l(\mu)\alpha}.\label{eq:cond_moment3}
\end{align}
In the same way as Section \ref{sec:JointFacPartitionSizes}, we factorize $(\theta+l(\mu)\alpha)^{l\uparrow\alpha}$ into $(\theta+l(\mu)\alpha)^{k\uparrow\alpha}$ and $(\theta+(l(\mu)+k)\alpha)^{l-k\uparrow\alpha}$ of which the latter stays inside the sum as a $l-k$th coefficient of the exponential generating function $(1-\alpha t)^{-\frac{\theta}{\alpha}-l(\mu)-k}$. Then, we can establish the FTRA representation of the sum in \eqref{eq:cond_moment3} as follows:
\begin{align}
    \sum_{l-k=0}^{m-s}\Big[\frac{t^{l-k}}{(l-k)!}\Big](1-\alpha t)^{-\frac{\theta}{\alpha}-l(\mu)-k}\Big[\frac{t^{m-s}}{(m-s)!}\Big](1-t)^{-n+l(\mu)\alpha}\frac{\big(\frac{1-(1-t)^\alpha}{\alpha}\big)^{l-k}}{(l-k)!} = (\theta+n+k\alpha)^{m-s\uparrow}\nonumber
\end{align}
Hence the target quantity is
\begin{align}
    \mathbb{E}_{n+m}\Big[\prod_{j=1}^m\rnew_j(\lambda\setminus\mu)^{k_j\downarrow}\Bigr|\mu\Big]  
    &=\frac{m!\theta^{n\uparrow}(\theta+l(\mu)\alpha)^{k\uparrow\alpha}(\theta+n+k\alpha)^{m-s\uparrow}}{(m-s)!\theta^{n+m\uparrow}}\prod_{j=1}^m\Big(\frac{(1-\alpha)^{j-1\uparrow}}{j!}\Big)^{k_j}
\end{align}
which coincides with the estimator given in \cite{cerquetti2013marginals}. Again, although some intermediate steps require careful treatment, the overall procedure is the same as in Sections \ref{sec:MomentPartitionLength} and \ref{sec:JointFacPartitionSizes} which showcases the uniformity of the approach.

\section{Conclusion and Discussion}\label{sec:Conclusion}
In this paper, Ewens-Pitman's partition structure is treated as an example of a harmonic function on a branching graph, specifically the Kingman graph. Then, we showed that the method of constructing this distribution via the interpolation polynomial approach proposed in \cite{borodin1999harmonic} amounts to carrying out umbral interpolation with Sheffer polynomial sequences. As a result, a new algebraic combinatorial representation of Ewens-Pitman's partition structure based on factorial monomial symmetric functions is obtained. This representation explicitly shows the decomposition between two sufficient statistics -- partition sizes and partition length -- an important characteristic of this distribution. Furthermore, we utilized the isomorphism between the Sheffer group and the Riordan group to show that the marginal of the Ewens-Pitman's partition structure taken with respect to all partitions of $n\in\mathbb{N}$ of fixed length has an FTRA representation. We used this result to obtain alternative ways to construct various summary statistics and estimators derived from the Ewens-Pitman's partition structure commonly seen in the literature. Our new approach consists primarily of symbolic methods of generating functions heavily utilizing the FTRA, thereby modernizing the tools typically used in the statistical literature surrounding these areas such as species sampling models in Bayesian nonparametrics.

These symbolic methods of obtaining various marginals of the Ewens-Pitman's partition structure are amenable to symbolic computation. In the future, efficient symbolic implementation of the methodology developed here is of great interest to the author. Furthermore, in this work, only the basic characteristics of the Riordan arrays were used. For example, Riordan arrays can also be characterized by a pair of power series that enable automatic derivation of recursion between elements in successive rows of a given Riordan array. Therefore, one may be able to use this property to devise an algorithm that automatically computes the recurrence relation of these various summary statistics and estimators we obtained. Last but not least, the interpolation polynomial approach can be applied to any branching graph provided that the symmetric function used to construct the harmonic function has the interpolation property. Therefore, the branching of other random combinatorial structures may also allow for similar representation and characterization introduced in this work.

\section*{Acknowledgement}
The author would like to thank Sylvia Fr{\"u}hwirth-Schnatter for careful reading of the draft and encouragement.
\begin{appendices}
\section{Derivations of \eqref{eq:FTRAinner_5}}\label{app:appA}
\begin{align}
    S_{n,l}^{-1,-\alpha,0}\frac{(-\alpha)^l l!}{j!}\sum_{i=0}^j(-1)^i\binom{j}{i}\frac{(l+i)^{i\downarrow}}{i!} &= S_{n,l}^{-1,-\alpha,0} \frac{(-\alpha)^l l!}{j!}\sum_{i=0}^j\Big[\frac{t^j}{j!}\Big]e^t\frac{(-t)^i}{i!}\Big[\frac{t^i}{i!}\Big]{}_{1}{F}_1(1+l,1,t)\label{eq:AppendixA_1}\\
    &=S_{n,l}^{-1,-\alpha,0} \frac{(-\alpha)^l l!}{j!}\Big[\frac{t^j}{j!}\Big]e^t{}_{1}{F}_1(1+l,1,-t)\label{eq:AppendixA_2}\\
     &=S_{n,l}^{-1,-\alpha,0} (-\alpha)^l[t^j]L_l(t)!\label{eq:AppendixA_3}\\
     &=S_{n,l}^{-1,-\alpha,0} (-\alpha)^l\Big[\frac{t^l}{l!}\Big]\Big(\frac{-1}{t-1}\Big)\frac{\Big(\frac{t}{t-1}\Big)^j}{j!}\label{eq:AppendixA_4}.
\end{align}
From left to right of \eqref{eq:AppendixA_1}, we first used 
$$
\frac{(l+i)^{i\downarrow}}{i!} = \frac{(l+1)^{i\uparrow}}{1^{i\uparrow}} = \Big[\frac{t^i}{i!}\Big]{}_1F_1(1+l,1,t)
$$
where ${}_1F_1(1+l,1,t)$ is the confluent hypergeometric function of the first kind. In addition, $(-1)^i\binom{j}{i}$ is the $(j,i)$-th element of $\mathcal{R}_e[e^t,-t]$. Hence we have the summation in the right hand of \eqref{eq:AppendixA_1}. From \eqref{eq:AppendixA_1} to \eqref{eq:AppendixA_2} we used the FTRA. From \eqref{eq:AppendixA_2} to \eqref{eq:AppendixA_3}, we used the known relationship between $e^t$ and ${}_1F_1(a,b,t)$ where 
\begin{align}
    e^t{}_1F_1(a,b,-t) = {}_1F_1(b-a,b,t)\nonumber
\end{align}
and that the confluent hypergeometric function of the first kind of form ${}_1F_1(-l,1,t)$ is equal to the Laguerre polynomial $L_l(t)$. Such Laguerre polynomial has a Riordan array representation as mentioned in \citet[p.~169]{shapiro2022riordan} (note their definition of Laguerre polynomial $\bold{L}_l(t)$ is equal to $l!L_l(t)$)  as follows:
\begin{align}
    \Big[\frac{t^j}{j!}\Big]l!L_l(t) = \Big[\frac{t^l}{l!}\Big]\Big(\frac{-1}{t-1}\Big)\frac{\Big(\frac{t}{t-1}\Big)^j}{j!},\nonumber
\end{align}
resulting in \eqref{eq:AppendixA_4} which is \eqref{eq:FTRAinner_5}.

\section{Proof for the new representation of $f_\lambda(\theta;\alpha)$}\label{appendix:B}
Here we show 
\begin{align}
    \sum_{i=0}^{l-j}(-1)^j\binom{i+j}{j}\frac{(\theta+l\alpha)^{i+j\uparrow\alpha}}{\alpha^{i+j}(i+j)!} = (-1)^j\frac{(\theta+l\alpha)^{l+1\uparrow\alpha}}{(\theta+(l+j)\alpha)\alpha^l(l-j)!j!}\label{eq:AppendixB_GOAL}
\end{align}
in Section \ref{sec:altrep} with $l=l(\lambda)$ using mathematical induction. First, when $j=0$, we have
\begin{align}
    \sum_{i=0}^l\frac{(\theta+l\alpha)^{i\uparrow\alpha}}{\alpha^{i}i!}
    &= \sum_{i=0}^l\nabla_\alpha^{l-i}\frac{(\theta+l\alpha)^{l\uparrow\alpha}}{\alpha^{l}l!}\label{eq:AppendixB_1}\\
    &=(\nabla_\alpha^l + \nabla_\alpha^{l-1}+\cdots + \nabla_\alpha+1)\frac{(\theta+l\alpha)^{l\uparrow\alpha}}{\alpha^{l}l!}\label{eq:AppendixB_2}
\end{align}
where in \eqref{eq:AppendixB_1} we used the operation of $\nabla_\alpha$ on ascending factorials defined as $\nabla_\alpha (\theta+l\alpha)^{k\uparrow\alpha} = k\alpha(\theta+l\alpha)^{k-1\uparrow\alpha},k=1,2,\ldots$. Now, as $\nabla^{l+i}_\alpha(\theta+l\alpha)^{l\uparrow\alpha} = 0$ for $i=1,2,\ldots$, we can extend the finite geometric sum of $\nabla_\alpha$ in \eqref{eq:AppendixB_2} to infinity and use the definition $\nabla_\alpha = 1-E^{-\alpha}$ to obtain the following:
\begin{align}
    \sum_{i=0}^\infty \nabla^i_\alpha\frac{(\theta+l\alpha)^{l\uparrow\alpha}}{\alpha^{l}l!} &= \frac{1}{1-\nabla_\alpha}\frac{(\theta+l\alpha)^{l\uparrow\alpha}}{\alpha^{l}l!} \nonumber\\
    &= \frac{1}{E^{-\alpha}}\frac{(\theta+l\alpha)^{l\uparrow\alpha}}{\alpha^{l}l!}\nonumber\\
    &=\frac{(\theta+(l+1)\alpha)^{l\uparrow\alpha}}{\alpha^ll!}\nonumber
\end{align}
which is equal to the right-hand side of \eqref{eq:AppendixB_GOAL} when $j=0$. Now, assuming that when $j=k$, the following holds,
\begin{align}
    \sum_{i=0}^{l-k}(-1)^k\binom{i+k}{k}\frac{(\theta+l\alpha)^{i+k\uparrow\alpha}}{\alpha^{i+k}(i+k)!} &= \sum_{i=0}^{l-k}(-1)^k\binom{i+k}{k}\nabla_\alpha^{l-k-i}\frac{(\theta+l\alpha)^{l\uparrow\alpha}}{\alpha^ll!}\label{eq:AppendixB_3}\\
    &=(-1)^k\frac{(\theta+l\alpha)^{l+1\uparrow\alpha}}{(\theta+(l+k)\alpha)\alpha^l(l-k)!k!}.\label{eq:AppendixB_4}
\end{align}
Comparing the right hand side of \eqref{eq:AppendixB_3} and \eqref{eq:AppendixB_4}, we can rewrite the right hand side of \eqref{eq:AppendixB_3} without the term $(-1)^k$  as follows:
\begin{align}
    \sum_{i=0}^{l-k}\binom{i+k}{k}\nabla_\alpha^{l-k-i}\frac{(\theta+l\alpha)^{l\uparrow\alpha}}{\alpha^ll!}= \frac{\theta+2l\alpha}{\theta+(l+k)\alpha}\binom{l}{k}\frac{(\theta+l\alpha)^{l\uparrow\alpha}}{\alpha^ll!}.\label{eq:AppendixB_whenjk}
\end{align}
We use this result to show the case when $j=k+1$ in \eqref{eq:AppendixB_GOAL}. First, the target quentity \eqref{eq:AppendixB_GOAL} when $j=k+1$ can be written as follows:
\begin{align}
    \sum_{i=0}^{l-k-1}(-1)^{k+1}\binom{i+k+1}{k+1}\frac{(\theta+l\alpha)^{i+k+1\uparrow\alpha}}{\alpha^{i+k+1}(i+k+1)!} &= (-1)^{k+1}\sum_{i=0}^{l-k-1}\binom{i+k+1}{k+1}\nabla_\alpha^{l-k-1-i}\frac{(\theta+l\alpha)^{l\uparrow\alpha}}{\alpha^ll!}\label{eq:AppendixB_5_2}\\
    &=\frac{(-1)^{k+1}}{\nabla_\alpha}\sum_{i=0}^{l-k-1}\binom{i+k+1}{k+1}\nabla_\alpha^{l-k-i}\frac{(\theta+l\alpha)^{l\uparrow\alpha}}{\alpha^ll!}\label{eq:AppendixB_5}.
\end{align}
Then we extend the range of summation in \eqref{eq:AppendixB_5} to $i= l-k$ and subtract the value at $i=l-k$ and obtain the following:
\begin{align}
    \sum_{i=0}^{l-k-1}\binom{i+k+1}{k+1}\nabla_\alpha^{l-k-i}\frac{(\theta+l\alpha)^{l\uparrow\alpha}}{\alpha^ll!} 
    &= \bigg(\sum_{i=0}^{l-k}\binom{i+k+1}{k+1}\nabla_\alpha^{l-k-i} - \binom{l+1}{k+1}\bigg)\frac{(\theta+l\alpha)^{l\uparrow\alpha}}{\alpha^ll!}\label{eq:AppendixB_6}.
\end{align}
Now, by using the recursion of the binomial coefficient, we have the following 
\begin{align}
    \sum_{i=0}^{l-k}\binom{i+k+1}{k+1}\nabla_\alpha^{l-k-i} = \sum_{i=0}^{l-k}\binom{i+k}{k}\nabla_\alpha^{l-k-i} + \sum_{i=0}^{l-k-1}\binom{i+k+1}{k+1}\nabla_\alpha^{l-k-i-1}\label{eq:AppendixB_7}.
\end{align}
The first sum of the right hand side of \eqref{eq:AppendixB_7} is the formula \eqref{eq:AppendixB_whenjk} without the term $\frac{(\theta+l\alpha)^{l\uparrow\alpha}}{\alpha^ll!}$ when $j=k$ while the second sum is when $j=k+1$. To summarize, \eqref{eq:AppendixB_5} can be written as follows:
\begin{align}
    \frac{(-1)^{k+1}}{\nabla_\alpha}\bigg(\sum_{i=0}^{l-k-1}\binom{i+k+1}{k+1}\nabla_\alpha^{l-k-i-1}+\sum_{i=0}^{l-k}\binom{i+k}{k}\nabla_\alpha^{l-k-i}-\binom{l+1}{k+1}\bigg)\frac{(\theta+l\alpha)^{l\uparrow\alpha}}{\alpha^ll!}.\label{eq:AppendixB_8}
\end{align}
which is equal to the target quantity \eqref{eq:AppendixB_5_2}. Combining and canceling terms, we obtain the following equation 
\begin{align}
   \frac{\nabla_\alpha-1}{\nabla_\alpha} \sum_{i=0}^{l-k-1}\binom{i+k+1}{k+1}\nabla_\alpha^{l-k-i-1}\frac{(\theta+l\alpha)^{l\uparrow\alpha}}{\alpha^ll!} = \frac{1}{\nabla_\alpha}\bigg(\sum_{i=0}^{l-k}\binom{i+k}{k}\nabla_\alpha^{l-k-i}-\binom{l+1}{k+1}\bigg)\frac{(\theta+l\alpha)^{l\uparrow\alpha}}{\alpha^ll!}.\label{eq:AppendixB_9}
\end{align}
By using the relation $\nabla_\alpha -1 = -E^{-\alpha}$ between the backward difference operator $\nabla_\alpha$ and the forward shift operator $E$, the left hand side quantity $\sum_{i=0}^{l-k-1}\binom{i+k}{k}\nabla_\alpha^{l-k-i-1}\frac{(\theta+l\alpha)^{l\uparrow\alpha}}{\alpha^ll!}$ in \eqref{eq:AppendixB_9} can be written as follows:
\begin{align}
    \sum_{i=0}^{l-k-1}\binom{i+k+1}{k+1}\nabla_\alpha^{l-k-i-1}\frac{(\theta+l\alpha)^{l\uparrow\alpha}}{\alpha^ll!} = E^\alpha\Bigg(\bigg(\binom{l+1}{k+1}-\sum_{i=0}^{l-k}\binom{i+k}{k}\nabla_\alpha^{l-k-i}\bigg)\frac{(\theta+l\alpha)^{l\uparrow\alpha}}{\alpha^ll!}\Bigg).\label{eq:AppendixB_10}
\end{align}
As mentioned before, the sum in the right hand side of \eqref{eq:AppendixB_10} times $\frac{(\theta+l\alpha)^{l\uparrow\alpha}}{\alpha^ll!}$ is the formula \eqref{eq:AppendixB_whenjk} when $j=k$ which we assume holds true to show the target quantity when $j=k+1$. Hence, we obtain the following
\begin{align}
    \sum_{i=0}^{l-k-1}\binom{i+k+1}{k+1}\nabla_\alpha^{l-k-i-1}\frac{(\theta+l\alpha)^{l\uparrow\alpha}}{\alpha^ll!} &= E^\alpha\Bigg(\bigg(\binom{l+1}{k+1}-\frac{\theta+2l\alpha}{\theta+(l+k)\alpha}\binom{l}{k}\bigg)\frac{(\theta+l\alpha)^{l\uparrow\alpha}}{\alpha^ll!}\Bigg)\label{eq:AppendixB_11}\\
    &=\bigg(\binom{l}{k+1}+\binom{l}{k}-\frac{\theta+(2l+1)l\alpha}{\theta+(l+k+1)\alpha}\binom{l}{k}\bigg)\frac{(\theta+(l+1)\alpha)^{l\uparrow\alpha}}{\alpha^ll!}\label{eq:AppendixB_12}\\
    &=\bigg(\binom{l}{k+1}-\frac{-(l+k)\alpha}{\theta+(l+k+1)\alpha}\frac{l!}{(l-k)!k!}\bigg)\frac{(\theta+(l+1)\alpha)^{l\uparrow\alpha}}{\alpha^ll!}\nonumber\\
    &=\bigg(\binom{l}{k+1}-\frac{-(k+1)\alpha}{\theta+(l+k+1)\alpha}\frac{l!}{(l-k-1)!(k+1)!}\bigg)\frac{(\theta+(l+1)\alpha)^{l\uparrow\alpha}}{\alpha^ll!}\nonumber\\
    &=\frac{\theta+l\alpha}{\theta+(l+k+1)\alpha}\binom{l}{k+1}\frac{(\theta+(l+1)\alpha)^{l\uparrow\alpha}}{\alpha^ll!}\nonumber\\
    &=\frac{(\theta+l\alpha)^{l+1\uparrow\alpha}}{(\theta+(l+k+1)\alpha)\alpha^l(l-k-1)!(k+1)!}.\label{eq:AppendixB_13}
\end{align}
From \eqref{eq:AppendixB_11} to \eqref{eq:AppendixB_12}, we used the property $E(f(\theta)g(\theta)) =  Ef(\theta) Eg(\theta) =f(\theta+1)g(\theta+1)$ of the shift operator $E$ \citep{jordan1965calculus} in addition to the recursion of the binomial coefficient. From the end result in \eqref{eq:AppendixB_13}, we establish the following:
\begin{align}
    \sum_{i=0}^{l-k-1}(-1)^{k+1}\binom{i+k+1}{k+1}\nabla_\alpha^{l-k-i-1}\frac{(\theta+l\alpha)^{l\uparrow\alpha}}{\alpha^ll!} = (-1)^{k+1}\frac{(\theta+l\alpha)^{l+1\uparrow\alpha}}{(\theta+(l+k+1)\alpha)\alpha^l(l-k-1)!(k+1)!}
\end{align}
which is \eqref{eq:AppendixB_GOAL} when $j=k+1$. Hence it holds for all $j\in\mathbb{N}$.
\section{Details of the derivations of joint conditional factorial moments of partitions sizes}\label{appendix:cond_details}
To simplify Equation \eqref{eq:cond_moment}, we first note that due to the conditioning of the paths, marginalization is taken with respect to $m$ additional observations. Furthermore, due to $\lambda$ being a disjoint union as given in \eqref{eq:LambdaDisjoint}, marginalization with respect to partition sizes (with fixed $l^\text{new}$) will take place separately. One with respect to $(q_1,\ldots,q_{l(\mu)})$ nonnegative increments on $(\mu_1,\ldots,\mu_{l(\mu)})$ with the total of $m-r$ observations allocated to them, the other with respect to $(1^{\rnew_1(\lambda\setminus\mu)}\ldots,r^{\rnew_{r}(\lambda\setminus\mu)})$ with the total of $r$ observations and we let such $r$ range between $0$ to $m$. Then, marginalization with respect to $l^\text{new}$ between 0 to $m$ will follow.

Now, from the result in Sections \ref{sec:UmbralInterpolationFTRA} and \ref{sec:altrep}, parts in $(-1)^{n+m}\dim(\lambda\setminus\mu)f_\lambda(\theta;\alpha)$ when marginalized with respect to partition sizes with fixed $l^\text{new}$ will result in a quantity indexed by partition sizes (e.g., generalized Strling numbers) is 
\begin{align}
    \frac{1}{\prod_{j=1}^{n+m}\rold_j(\lambda)!\prod_{j=1}^m\rnew_j(\lambda\setminus\mu)!}\prod_{j=1}^m\Big(\frac{(1-\alpha)^{j-1\uparrow}}{j!}\Big)^{\rnew_j(\lambda\setminus\mu)}\prod_{j=1}^{l(\mu)}(1-\alpha)^{\mu_j+q_j-1\uparrow}\label{eq:appc_1}
\end{align}
of $(-1)^{n+m}f_\lambda(\theta;\alpha)$ and $\dim(\lambda\setminus\mu)$. Based on this understanding, we carry out the marginalization with respect to $\rnew$ for \eqref{eq:appc_1} and $\dim(\lambda\setminus\mu)$: 
\begin{align}
    &\sum_{r=0}^m\sum_{\substack{\rnew=r\\l^\text{new}=l}}\dim(\lambda\setminus\mu)\frac{1}{\prod_{j=1}^{n+m}\rold_j(\lambda)!\prod_{j=1}^m\rnew_j(\lambda\setminus\mu)!}\prod_{j=1}^m\Big(\frac{(1-\alpha)^{j-1\uparrow}}{j!}\Big)^{\rnew_j(\lambda\setminus\mu)}\prod_{j=1}^{l(\mu)}(1-\alpha)^{\mu_j+q_j-1\uparrow}\nonumber\\
    &=\sum_{r=0}^m\Big(\sum_{\substack{\rnew=r\\l^\text{new}=l}}\prod_{j=1}^r\Big(\frac{(1-\alpha)^{j-1\uparrow}}{j!}\Big)^{\rnew_j(\lambda\setminus\mu)}\frac{r!}{\rnew_j(\lambda\setminus
    \mu)!}\Bigg)\Bigg(\sum_{\substack{q_1+\cdots+q_{l(\mu)}\\ = m+r}}\frac{\prod_{j=1}^{n+m}\rold_j(\lambda)!}{\prod_{j=1}^n\rold_j(\mu)!}\frac{(m-r)!}{\prod_{j=1}^{n+m}\rold_j(\lambda)!\prod_{j=1}^lq_j!}\prod_{j=1}^{l(\mu)}(1-\alpha)^{\mu_j+q_j-1\uparrow}\Bigg)\nonumber
\end{align}
where the factor $\frac{\prod_{j=1}^{n+m}\rold_j(\lambda)!}{\prod_{j=1}^n\rold_j(\mu)!}$ appears as the multiplicity of the increments $(q_1,\ldots,q_{l(\mu)})$. Namely,  $\prod_{j=1}^{n+m}\rold_j(\lambda)!$ as the multiplicity of old parts of $\lambda$ divided by $\prod_{j=1}^n\rold_j(\mu)!$ the multiplicity of parts of $\mu$. Thus terms $\prod_{j=1}^{n+m}\rold_j(\lambda)!$ will cancel with each other in the second big bracket. Dividing remaining terms in the second big bracket with the denominator $(-1)^nf_\mu(\theta;\alpha)$ in \eqref{eq:cond_moment} (as $\dim(\mu)$ cancels with the first term in the factorization $\dim(\lambda) = \dim(\mu)\dim(\lambda\setminus\mu)$) results in the following:
\begin{align}
   & \Bigg(\sum_{\substack{q_1+\cdots+q_{l(\mu)}\\ = m+r}}\frac{(m-r)!}{\prod_{j=1}^n\rold_j(\mu)!\prod_{j=1}^lq_j!}\prod_{j=1}^{l(\mu)}(1-\alpha)^{\mu_j+q_j-1\uparrow}\Bigg)/\Bigg(\frac{\theta^{l(\mu)\uparrow\alpha}}{\prod_{j=1}^n\rold_j(\mu)}\prod_{j=1}^{l(\mu)}(1-\alpha)^{\mu_j-1\uparrow}\Bigg)\\
   &=\frac{1}{\theta^{l(\mu)\uparrow\alpha}}\Bigg(
   \sum_{\substack{q_1+\cdots+q_{l(\mu)}\\ = m+r}}
   \binom{m-r}{q_1,\ldots,q_{l(\mu)}}\prod_{j=1}^{l(\mu)}(\mu_j-\alpha)^{q_j\uparrow}\Bigg)
\end{align}
Marginalization with respect to partition sizes $\rnew$ with fixed partition length $l^\text{new}= l$ for $(-1)^{n+m}\dim(\lambda\setminus\mu)f_\lambda(\theta;\alpha)$ for parts except \eqref{eq:appc_1} will result in $\theta^{l^\text{new}+l(\mu)\uparrow\alpha}$ as one can easily check from results in Section \eqref{sec:UmbralInterpolationFTRA}. Combining all these results plus the use of the derivative rule to replace $\prod_{j=1}^m\rnew_j(\lambda\setminus\mu)^{k_j\downarrow}$ with $\prod_{j=1}^m(\xi_jD_{\xi_j})^{k_j}$ for $\xi_j =  \frac{(1-\alpha)^{j-1\uparrow}}{j!}$ results in Equation \eqref{eq:cond_moment2}.
\end{appendices}
\bibliography{references}  


%
%
%
%

\end{document}